\documentclass[fleqn,usenatbib,dvipsnames]{mnras}

\usepackage{newtxtext,newtxmath}
\usepackage{multirow}
\usepackage[T1]{fontenc}
\usepackage[version=4]{mhchem}
\usepackage{orcidlink}

\DeclareRobustCommand{\VAN}[3]{#2}
\let\VANthebibliography\thebibliography
\def\thebibliography{\DeclareRobustCommand{\VAN}[3]{##3}\VANthebibliography}

\usepackage{graphicx}
\usepackage{amsmath}
\usepackage{physics}
\usepackage{xcolor}
\usepackage{soul}

\soulregister\citep7
\soulregister\citet7
\soulregister\citealt7
\soulregister\ref7

\title[$\beta$-Particle Thermalization in Kilonova Ejecta]{Beta-Particle Transport and Thermalization in Kilonova Ejecta with Detailed Atomic Microphysics}

\author[Z. L. Andalman et al.]{
Zachary L. Andalman\,\orcidlink{0000-0001-5064-1269}$^{1, 2}$\thanks{E-mail: zack.andalman@princeton.edu},
Christopher L. Fryer\,\orcidlink{0000-0003-2624-0056}$^{2,3,4}$,
Christopher J. Fontes\,\orcidlink{0000-0003-1087-2964}$^{2,5}$,
\newauthor 
Matthew R. Mumpower\,\orcidlink{0000-0002-9950-9688}$^{7,2,6,8}$,
and Ryan T. Wollaeger\,\orcidlink{0000-0003-3265-4079}$^{2,6}$
\\
$^{1}$Department of Astrophysical Sciences, Princeton University, 4 Ivy Lane, 08540, Princeton, NJ, USA\\
$^{2}$Center for Theoretical Astrophysics, Los Alamos National Laboratory, Los Alamos, NM 87545, USA\\
$^{3}$Theoretical Division, Los Alamos National Laboratory, Los Alamos, NM 87545, USA\\
$^{4}$Department of Physics, The George Washington University, Washington, DC 20052, USA \\
$^{5}$Computational Physics Division, Los Alamos National Laboratory, Los Alamos, NM 87545, USA\\
$^{6}$Computing and Artificial Intelligence Division, Los Alamos National Laboratory, Los Alamos, NM 87545, USA\\
$^{7}$Obsidian Research, Fort Wayne, IN 46835, USA\\
$^{8}$Department of Physics and Astronomy, University of Notre Dame, Notre Dame, IN, 46656, USA
}

\date{Accepted XXX. Received YYY; in original form ZZZ}

\pubyear{2026}

\begin{document}
\label{firstpage}
\pagerange{\pageref{firstpage}--\pageref{lastpage}}
\maketitle

\begin{abstract}
When two neutron stars collide, they eject material containing heavy nuclei formed by the rapid neutron capture process ($r$-process). As these nuclei decay, they power a bright optical/near-infrared transient known as a kilonova (KN). Modeling KN emission is a complex problem involving atomic opacities, radiation transport, and heating powered by the thermalization of radioactive decay products like $\gamma$-rays, $\alpha$-particles, and $\beta$-particles. For heating by $\gamma$-rays, many KN modeling codes do full radiation transport calculations. However, heating by $\alpha$- and $\beta$-particles relies on simplified descriptions of collisions and transport, and remains an important source of uncertainty in KN models. In this paper, we study the thermalization and transport of $\beta$-particles. To study thermalization, we use evaluated atomic physics data to estimate per-species contributions to energy deposition, scattering, and electron impact ionization, which we make available online. To include non-local effects, we develop a fully relativistic framework for charged particle transport in a spherically symmetric, homologously expanding ejecta, considering two limiting magnetic-field geometries. Non-local energy deposition and escape reduce thermalization efficiency, especially in the innermost and outermost ejecta, lowering the ejecta temperature and ionization state compared to local deposition models. Coulomb scattering partially offsets these effects by trapping particles at intermediate times $\sim 3-30$ days post-merger. Ionization by secondary electrons significantly enhances the overall ionization rate. We provide analytic prescriptions for the spatially dependent thermalization efficiency for use in future light-curve calculations. Our results demonstrate that evaluated atomic data and charged-particle transport should be incorporated into the next generation of KN models.
\end{abstract}


\section{Introduction}
\label{sec:intro}
In binary neutron star (BNS) mergers and some black hole-neutron star mergers, the radioactive decay of heavy nuclides produced by the rapid neutron capture process (\textit{r}-process) powers a bright flare in the optical and near-infrared (IR) known as a kilonova (KN) \citep{Li&Paczynski1998, Metzger+2010}, accompanied by a short gamma ray burst (GRB) \citep{Paczynski1986, Eichler+1989, Narayan+1992} and gravitational waves (GWs). The first confirmed KN, the bright optical transient AT 2017gfo, was discovered in association with gravitational wave event GW170817 \citep{Abbott+2017a} and the short GRB 170817A \citep{Goldstein+2017, Savchenko+2017}, providing compelling evidence that BNS mergers are sites of $r$-process nucleosynthesis \citep{Pian+2017, Tanvir+2017}.

The multi-band, multi-messenger nature of these events makes them a powerful astrophysical tool for constraining cosmological parameters \citep{Nissanke+2013b, Abbott+2017b}, binary star evolution models \citep{Fong&Berger2013, Fong+2022}, cosmic nucleosynthesis models \citep{Kasen+2017, Cote+2018, Rosswog+2018}, the physics of the $r$-process, and the equation of state for dense nuclear matter \citep{Bauswein+2013, Margalit&Metzger2017}. 

The $r$-process is expected to operate in material ejected from the system dynamically \citep{Lattimer&Schramm1974, Freiburghaus+1999, Rosswog+1999, Goriely+2011, Korobkin+2012} or unbound from the remnant accretion disk by neutrino-driven \citep{Dessart+2009, Perego+2014, Radice+2018} or viscously-driven \citep{Nedora+2019, Nedora+2021} winds \citep[for a review, see][]{Rosswog&Korobkin2024}. The ejecta are heated and ionized by the thermalization of radioactive decay products like $\beta$-particles, $\alpha$-particles, $\gamma$-rays, and fission fragments \citep{Barnes+2016}. 

The observed light curves and spectra depend on the ejecta density and velocity structure, ejecta morphology, wavelength-dependent opacity, and radioactive heating rate \citep{Metzger+2010, Roberts+2011, Kasen+2017, Tanaka+2017, Wollaeger+2018, Bulla2019, Even+2020, Hotokezaka&Nakar2020, Kawaguchi+2020, Korobkin+2021, Wollaeger+2021, Zhu+2021, Fryer+2024}. Both the opacity and heating rate are sensitive to nuclear physics models \citep{Barnes+2021}, ejecta composition, and the efficiency of decay product thermalization, since the decay energy may escape or be lost by adiabatic expansion \citep{Barnes+2016, Kasen&Barnes2019}. 

The bound-bound opacity is also sensitive to ejecta composition, especially of lanthanide elements which have an outsized contribution due to their valence $f$-shell electrons \citep{Barnes&Kasen2013, Kasen+2013, Tanaka&Hotokezaka2013, Wollaeger+2018, Fontes+2020, Fontes+2026}. The opacity is additionally sensitive to the ejecta ionization and excitation structure \citep{Pognan+2022b, Pognan+2023, Banerjee+2025, Pognan+2025}.

The thermalization efficiency is a key uncertainty because it depends on the interactions between radioactive decay products and the ejecta, for which the relevant atomic data are often unavailable despite ongoing work \citep{Kasen+2013, Fontes+2015, Fontes+2020, Pognan+2022}. Since decay products can travel significant distances in the ejecta before depositing their energy, thermalization also depends on charged-particle transport, and therefore on the magnetic field, whose geometry is largely unconstrained \citep{Barnes+2016}.

The ionization structure is subject to similar uncertainties. During the nebular phase, the ejecta are primarily ionized by non-thermal decay products \citep{Hotokezaka+2021, Pognan+2022, Brethauer+2026}, so the ionization rate depends both on thermalization and impact ionization cross sections. For the conditions in KN ejecta, recombination is often dominated by the atomic process known as dielectronic capture, whose rates are sensitive to fine-structure effects \citep{Hotokezaka+2021, Banerjee+2025, Singh+2025}. 

In the present work, we focus on $\beta$-particles, which carry most of the decay energy, and revisit their thermalization and transport. Building on previous thermalization efficiency calculations \citep{Barnes+2016, Hotokezaka+2016, Kasen&Barnes2019, Waxman+2019, Hotokezaka&Nakar2020, Shenhar+2024, VandenBerg&Hotokezaka2026}, we incorporate evaluated atomic physics data and estimate per-species contributions to energy deposition and electron impact ionization. We also estimate per-species contributions to scattering, whose role in particle transport has previously been neglected. We combine these data using abundance patterns generated from our nuclear network calculations, but we make our atomic data available online for use with arbitrary compositions (see the \hyperref[sec:data]{Data Availability statement}).

Despite the importance of thermalization, explicit charged-particle transport has only been considered by a few KN studies \citep{Barnes+2016, Barnes+2021, Wollaeger+2024}. We build on this work by developing a fully relativistic framework for particle transport in a spherically symmetric, homologously expanding ejecta, and by including an expanded set of microphysical phenomena. 

We consider two limiting transport scenarios: no transport, corresponding to highly tangled or toroidal magnetic fields, and maximal transport, corresponding to weak or coherent radial magnetic fields. The no transport scenario is equivalent to local deposition models considered by previous works \citep{Hotokezaka+2016, Kasen&Barnes2019, Waxman+2019, Hotokezaka&Nakar2020}. We investigate the effect of non-local energy deposition on the spatially dependent thermalization efficiency and ionization balance. We provide analytic prescriptions for the spatially dependent thermalization efficiency for use in future light curve calculations.

This paper is structured as follows. In Section~\ref{sec:microphysics}, we describe our approach to microphysical processes, including $r$-process yields, decay spectra, and electron-atom interactions. In Section~\ref{sec:global_model}, we describe our idealized global ejecta and particle transport model. In Section~\ref{sec:results}, we show the results of our global model calculations and explore their implications for thermalization. In Section~\ref{sec:ionstate}, we post-process our fiducial model to determine the temperature and ionization structure as an illustrative calculation. We conclude in Section~\ref{sec:conclusion}.

\section{Microphysics}
\label{sec:microphysics}
\subsection{Nuclear yields}

In the seconds after the BNS merger, the high temperature and neutron flux create the necessary conditions for $r$-process nucleosynthesis \citep{Miller2019}. Using the nuclear reaction network code \textsc{Prism} \citep{Sprouse+2021}, we calculate nuclear yields for three thermodynamic trajectories taken from tracer particles in 3D hydrodynamic simulations. Our strong $r$-process trajectory comes from a relativistic hydrodynamics simulation of a BNS merger in \citet{Goriely+2011} with an electron fraction $\simeq 0.1$, as expected for BNS merger dynamical ejecta. Our medium and weak $r$-process trajectories come from general relativistic neutrino radiation magnetohydrodynamics simulations of black hole accretion in \citet{Sprouse+2024} using electron fractions $\simeq 0.2$ and $0.4$ respectively, as expected for BNS merger disk winds. 

The nuclides produced by the $r$-process proceed towards the valley of stability via $\beta^-$-decay processes of the form
\begin{equation}
    \ce{^A_{Z}X} \to \ce{^{A}_{Z+1}X} + e^- + \overline{\nu}_e \,,
\end{equation}
where $Z$ is the atomic number, $A$ is the mass number, and $\ce{^A_{Z}X}$ is an isotope of element $X$. $\beta$ decays produce an electron anti-neutrino and a high-energy electron, or $\beta$-particle. If the daughter nuclide is in an excited nuclear state, it can additionally emit $\gamma$-rays or neutrons as it de-excites. 

\begin{figure}
    \centering
    \includegraphics[width=1.0\linewidth]{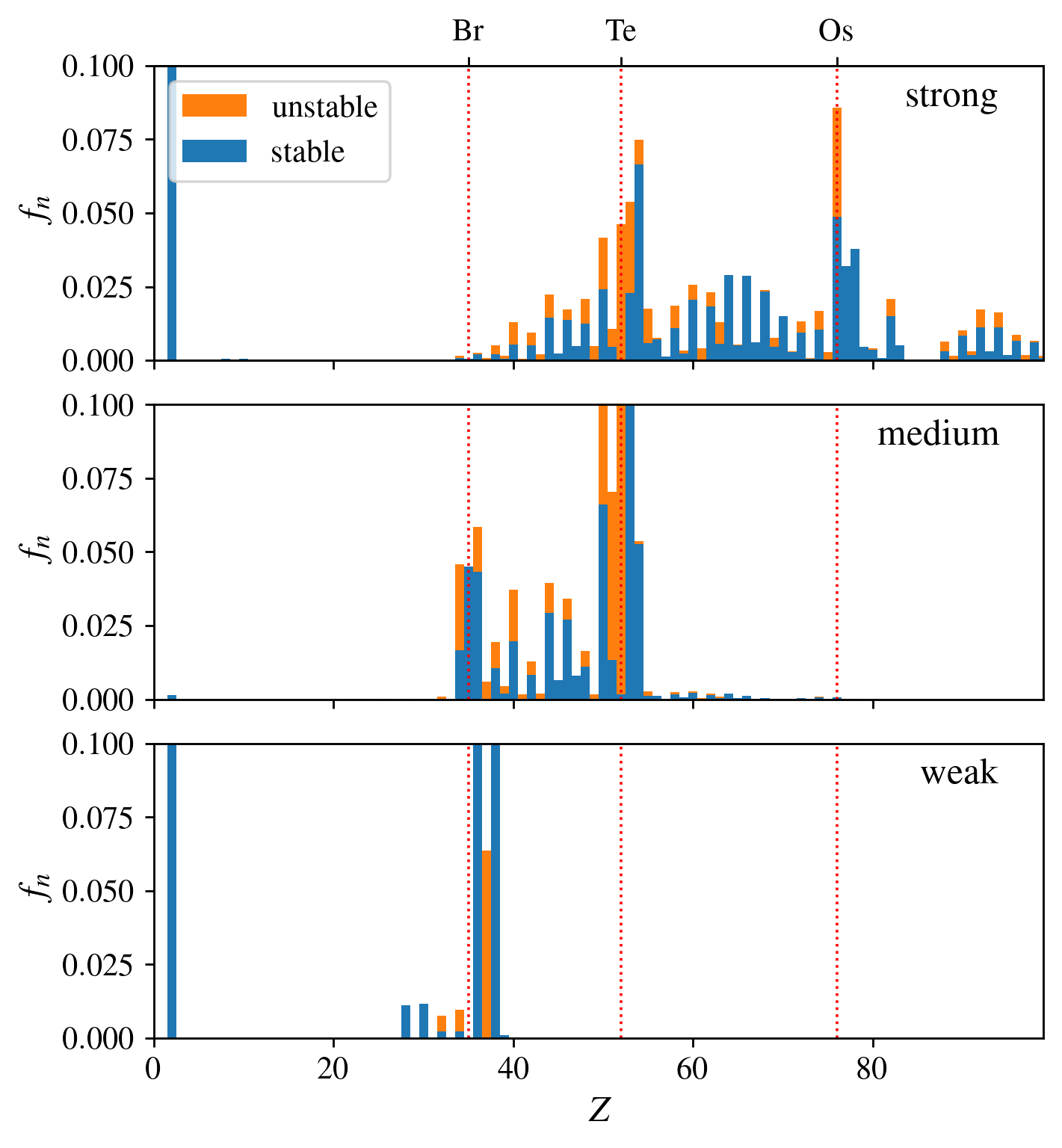}
    \caption{Elemental number fractions at $30~{\rm days}$ post-merger for our strong (top panel), medium (middle panel), and weak (bottom panel) $r$-processes. Each element is partitioned into stable (blue) and unstable (orange) isotopes. Bromine ($Z=35$), Tellurium ($Z=52$), and Osmium ($Z=76$) are marked by vertical, red-dotted lines as representatives of the $r$-process peaks. Our weak $r$-process yields only first-peak elements. Our medium $r$-process yields nuclei up to and around Te. Our strong $r$-process yields both third-peak nuclei and actinides.}
    \label{fig:ab}
\end{figure}

The $r$-process preferentially synthesizes nuclides near specific mass numbers $A \sim 80$, $130$, and $195$, known as $r$-process peaks, where neutron-rich nuclei exhibit enhanced stability due to closed nuclear shells. In Figure~\ref{fig:ab}, we plot the elemental abundances of our three $r$-processes at 30 days post-merger. Our weak $r$-process yields only first peak elements ($Z \sim 40$, $A \sim 80$, $\mu_{\rm mmw} \simeq 42$), where $\mu_{\rm mmw}$ is the mean molecular weight, ignoring free electrons. Our medium $r$-process yields nuclei up to and around Te ($Z = 52$, $A \sim 130$, $\mu_{\rm mmw} \simeq 115$). Our strong $r$-process yields both third peak nuclei and actinides.

Our strong and weak $r$-processes produce a significant number of $\alpha$ particles ($Z=2$). In the strong scenario, this is due to the production of $\alpha$ particles from the decay of actinide species \citep{Lund2023}. In the weak scenario, this is instead due to the conversion of neutrons and protons into $\alpha$ particles during the early stages of nucleosynthesis \citep{McLaughlin1996}. 

\citet{Feng+2026} recently showed that $r$-process yields may be sensitive to magnetically powered outbursts from long-lived merger remnants. Shocks increase the electron fraction by heating the ejecta, which accelerates weak reactions like positron capture on neutrons, suppressing 3rd-peak production. However, they also increase the ejecta entropy, enhancing heavy element production. The final abundances are therefore still highly uncertain, motivating our choice to consider a range of abundance patterns.

\subsection{$\beta$ decay spectra}

For each nuclide, we source $\beta$ decay rates and branching ratios from the International Atomic Energy Agency (IAEA) Livechart\footnote{\href{https://www-nds.iaea.org/relnsd/vcharthtml/VChartHTML.html}{www-nds.iaea.org/relnsd/vcharthtml/VChartHTML.html}}. Livechart data is directly sourced from the Evaluated Nuclear Structure Data File (ENSDF)\footnote{\href{https://www.nndc.bnl.gov/ensdf/}{www.nndc.bnl.gov/ensdf/}}, which is compiled from a review of available experimental data supplemented with systematic trend studies and theoretical models.

We compute $\beta$ decay spectra from the coupled Quasiparticle Random Phase Approximation and Hauser-Feshbach (QRPA+HF) model \citep{Mumpower2016, Moller2019, Mumpower+2025}. The model occasionally fails for nearly stable nuclei because there is little energy available for $\beta$ decay, so small errors in the nuclear level states can produce large errors in the decay spectrum. In these cases, we fall back to the $\beta$ decay spectra in Livechart, which are computed with the code \textsc{BetaShape} \citep{Mougeot2015a, Mougeot2015b}.

For each nuclide, the specific decay rate (units ${\rm s^{-1} g^{-1}}$) is
\begin{equation}
    \Gamma_{\rm dec} = N_{\rm A} (\ln 2) Y b \tau_{\rm h}^{-1} \,,
    \label{eq:dec_rate_nuc}
\end{equation}
where $N_{\rm A}$ is Avogadro's number, $Y$ is the isotopic abundance (units {\rm mol/g}), $b$ is the $\beta^-$ decay branching ratio, and $\tau_{\rm h}$ is the half-life. Most $r$-process nuclides have branching ratios of unity, but a handful have other significant decay pathways.

Inconsistencies between decay rates in \textsc{Prism}'s nuclear reaction network and our evaluated data sources used for post-processing could lead to errors in energy injection rates and $\beta$-decay spectra, as pointed out by \citet{Guttman+2024}. We test this possibility by comparing the decay rates from Equation~\ref{eq:dec_rate_nuc} to those computed directly from the \textsc{Prism} reaction flow. We find discrepancies $\lesssim 5\%$ in the injection rate and mean decay energy at early times $\lesssim 0.1~{\rm day}$. However, at later times $\gtrsim 1~{\rm day}$ relevant for our analysis, the discrepancies become negligible, except for our medium $r$-process for which percent-level discrepancies persist to late times. We plan to modify the reaction network to correct these in future work.

Let $\dd n^{(Z, A)}(E)$ denote the number of $\beta$-particles produced in an infinitesimal energy bin $\dd E$ by nuclide $(Z, A)$. These spectra are normalized to unity, the number of $\beta$ particles produced per decay, following nuclear physics conventions. The total decay spectrum (units ${\rm s^{-1} g^{-1} erg^{-1}}$) is the sum of contributions from each nuclide,
\begin{equation}
    \dd \lambda_{\rm dec}(E) = \sum_{Z,A} \Gamma_{\rm dec}^{(Z, A)} \dd n^{(Z, A)}(E) \,.
\end{equation}

For our strong $r$-process at $30~{\rm days}$ post-merger, there are 339 distinct unstable nuclei with nonzero abundances. We compute decay spectra for 287 nuclei and use the Livechart spectra for 40 nuclei. No spectra are available for the remaining 12 nuclei, but they produce a negligible fraction of the decay electrons ($< 10^{-6}$ of the decay rate), so they can be neglected.

\begin{figure}
    \centering
    \includegraphics[width=\linewidth]{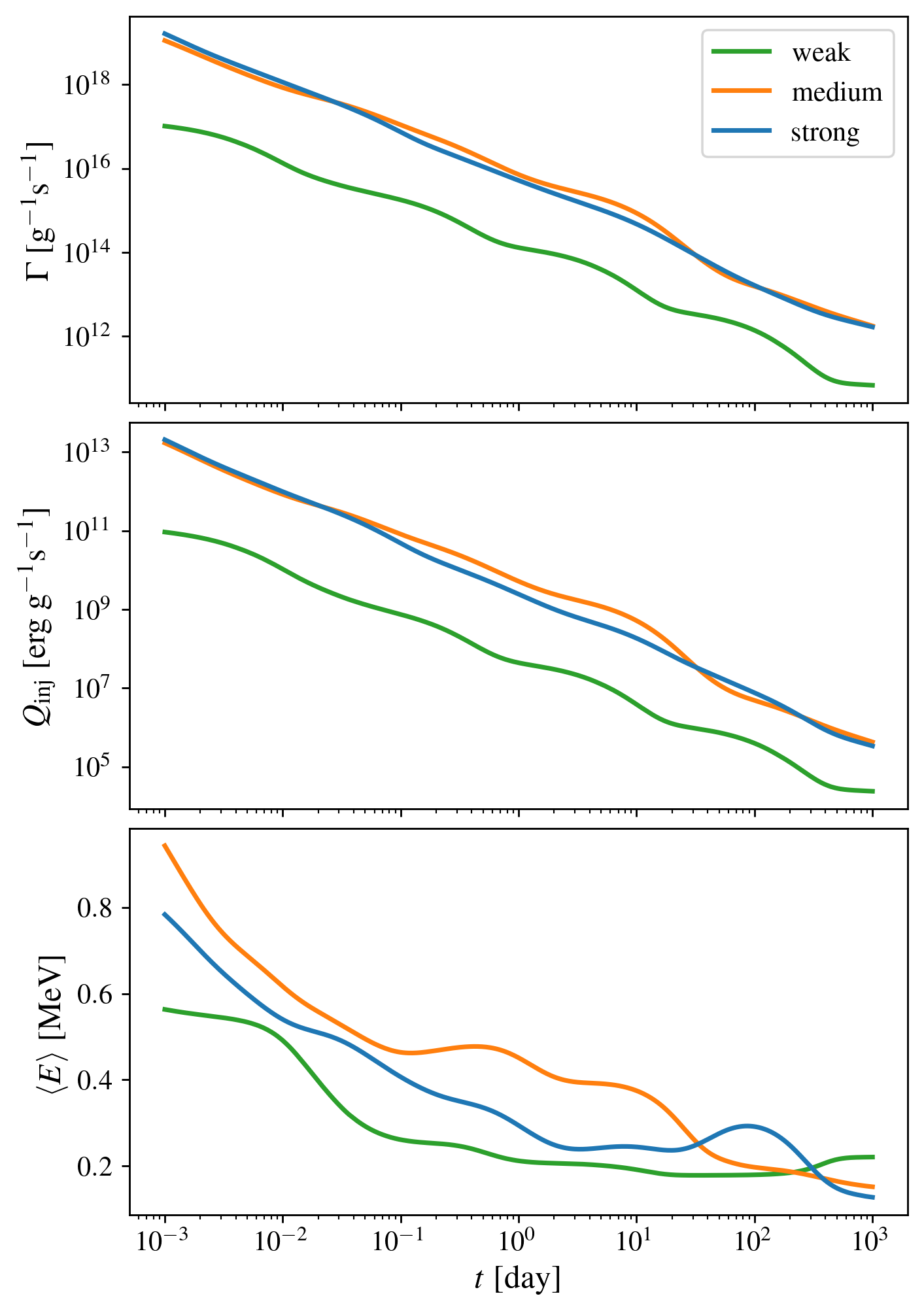}
    \caption{The $\beta$ decay rate (top panel), the $\beta$ energy injection rate (middle panel), and the mean $\beta$-particle energy (bottom panel) as a function of time for our strong (blue), medium (orange), and weak (green) $r$-processes. The dashed lines show the contribution from only the steady-state term (Eq.~\ref{eq:dec_rate_nuc}). The energy injection rates drop rapidly as nuclides reach the line of $\beta$ stability, and can be approximated by power laws.}
    \label{fig:rate}
\end{figure}

In Figure~\ref{fig:rate}, we plot the $\beta$ decay and energy injection rates and the mean $\beta$-particle energy for our three $r$-processes.  The energy injection rates drop rapidly as nuclides reach the line of $\beta$ stability, and can be approximated by the following power laws:
\begin{equation}\begin{split}
	Q_{\rm inj}^{({\rm strong})} \sim\ & 2.65 \times 10^{9} t_{\rm d}^{-1.29}~{\rm erg\ s^{-1} g^{-1}} \label{eq:inj_rate}\\
	Q_{\rm inj}^{({\rm medium})} \sim\ & 3.6 \times 10^{9} t_{\rm d}^{-1.27}~{\rm erg\ s^{-1} g^{-1}}\\
	Q_{\rm inj}^{({\rm weak})} \sim\ & 5.31 \times 10^{7} t_{\rm d}^{-1.14}~{\rm erg\ s^{-1} g^{-1}} \,,
\end{split}\end{equation}
where $t_{\rm d}$ is the time in days. We compute the power law prefactor and exponent using a least-squares linear regression in log space. For our strong $r$-process, the power law exponent is consistent with previous works, which find values ranging from $-1.4$ to $-1.1$ \citep{Metzger+2010, Goriely+2011, Roberts+2011, Korobkin+2012, Hotokezaka+2017}. 

The strong and medium $r$-processes have similar decay energies, but the weak $r$-process is less energetic by a factor $\sim 50$. The mean $\beta$-particle energy decreases gradually from around $0.8~{\rm MeV}$ at early times to around $0.2~{\rm MeV}$ at late times. 
The decreasing decay energy can be explained by the negative correlation between decay half-life and $Q$-values \citep{Sprouse+2022}, but it does not always hold due to the spread in decay properties and the impact of inverted decay chains, in which a decay with a long half-life produces a short-lived daughter nucleus \citep{Waxman+2019, Shenhar+2024}.

\begin{figure}
    \centering
    \includegraphics[width=\linewidth]{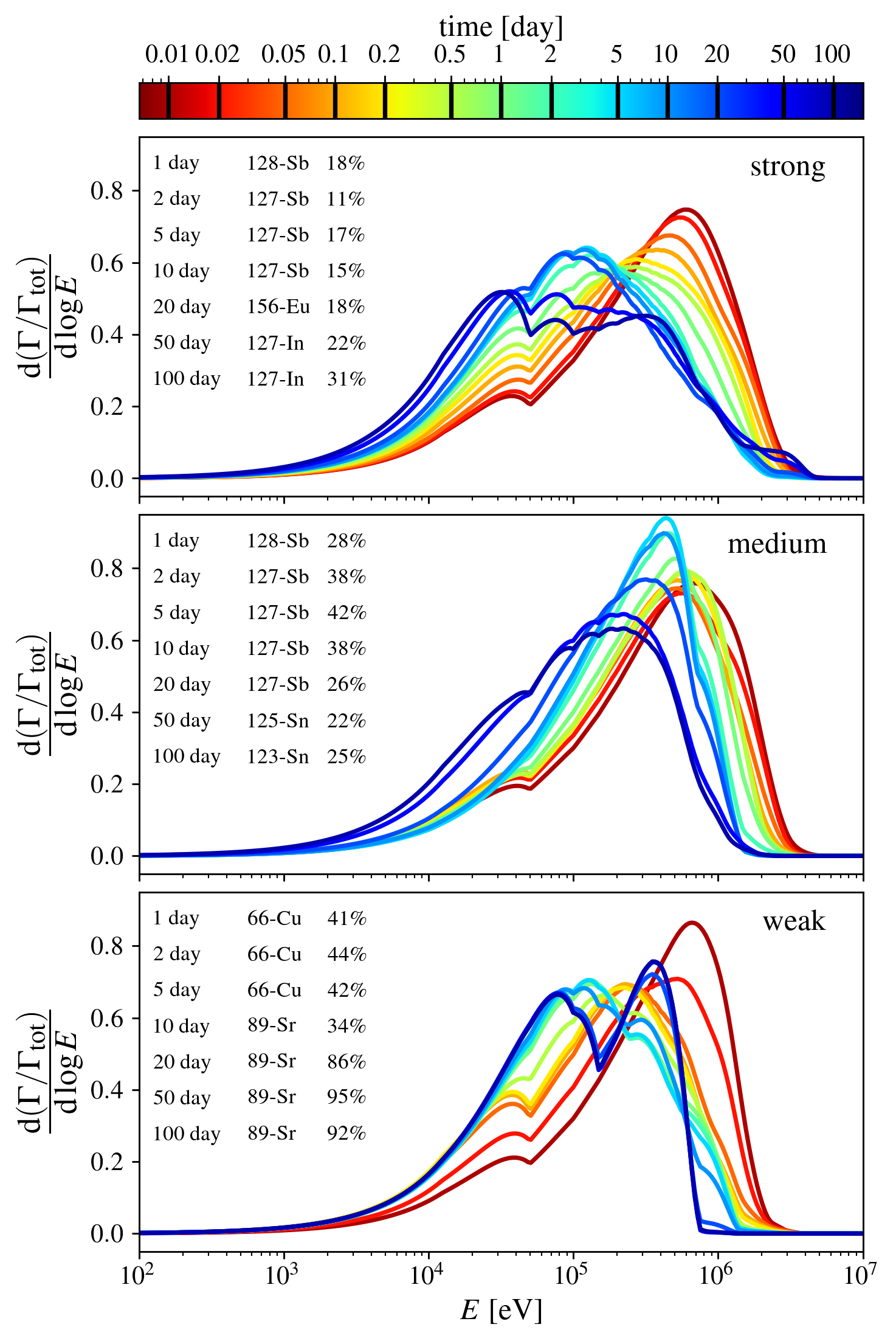}
    \caption{The $\beta$ decay spectra as a function of time for our strong (top panel), medium (middle panel), and weak (bottom panel) $r$-processes. Each time is shown in a different color indicated by the colorbar. At various times, we indicate the nuclide with the largest contribution to the energy injection rate and its fractional contribution to the energy. At $0.01$ days, the $\beta$ decay spectra peak at $\sim 1~{\rm MeV}$. Over time, the spectra shift to lower energies because higher-energy decays are generally associated with shorter-half-life nuclides. At each time, the spectra are dominated by contributions from only a handful of nuclides, producing lumpy features.}
    \label{fig:decspec}
\end{figure}

In Figure~\ref{fig:decspec}, we plot the normalized $\beta$ spectra for our three $r$-processes at various times. We also list the nuclide that contributes most to the decay at 1, 2, 5, 10, 20, 50, and 100 days post-merger. At $0.01$ days, the $\beta$ decay spectra peak at $\sim 1~{\rm MeV}$. Over time, the spectra shift to lower energies, consistent with Figure~\ref{fig:rate}. At each time, the spectra are dominated by contributions from only a handful of nuclides, producing lumpy features. 

We do not account for long-lived nuclear isomers of parent nuclides, such as those of $\ce{^{128}Sb}$, $\ce{^{85}Kr}$, $\ce{^{129}Te}$, and $\ce{^{131}Te}$, which may alter the timing of $\beta$-decay energy release and modify the spectrum \citep{Fujimoto&Hashimoto2020, Misch+2021a, Misch+2021b, Misch&Mumpower2024}. Specifically, these isomers can act either as `accelerants' that expedite energy release or as `batteries' that store it for later times. Isomeric transitions may also produce detectable $\gamma$-rays in next-generation telescopes \citep{Hamilton+2026}. Future work should investigate whether this effect is significant and, if so, under what conditions.

\subsection{Electron-atom interactions}
\label{sec:elec-atom_inter}

Suprathermal $\beta$-particles can undergo many interactions with the atoms of the background plasma, including 
\begin{enumerate}
    \item Electron impact ionization (EII): an energetic electron interacts with an atom and produces an ion and a secondary electron;
    \item Electron impact excitation (EIE): an energetic electron interacts with an atom and promotes the atom to an excited state;
    \item Electron-atom bremsstrahlung: an electron interacts with an atom via Coulomb forces, producing a photon;
    \item Mott scattering: an electron interacts with an atom via Coulomb forces, resulting in elastic scattering.
\end{enumerate}

We assume that scattering is dominated by (iv), consistent with the scattering theory and electron transport literature \citep{Fano1954, Turner&Hamm1995, Mayol&Salvat1997}, so we ignore scattering contributions from (i)--(iii). For energy loss, we work in the continuous slowing-down approximation (CSDA), approximating discrete electron-atom interactions as a continuous stopping cross section given by
\begin{equation}
	S(E) = \int \dd (\Delta E) \Delta E \dv{\sigma(E)}{\Delta E} \,,
	\label{eq:stop}
\end{equation}
where $\dd \sigma(E) / \dd \Delta E$ is a differential cross section in energy loss $\Delta E$. The corresponding energy loss rate is $-\dd E / \dd t = n v S(E)$, where $n$ is the target number density and $v$ is the $\beta$-particle velocity.

CSDA is a good approximation for EII and EIE, but less so for bremsstrahlung, where the energy loss is dominated by rare interactions that consume a significant fraction of the incident electron's kinetic energy. However, in the following sections we show that the bremsstrahlung contribution to the energy loss is subdominant at the typical $\beta$-particle energies.

For scattering (iv), the differential cross section diverges at small angles. Rather than modeling many small-angle scatters, we model scattering as isotropic using the transport (momentum-transfer) cross section given by
\begin{equation}
	\sigma_{\rm t} = \int \dd \mu_{\rm s} (1 - \mu_{\rm s}) \dv{\sigma(E)}{\mu_{\rm s}} \,,
	\label{eq:sigt}
\end{equation}
where $\dd \sigma(E) / \dd \mu_{\rm s}$ is a differential cross section in cosine scattering angle $\mu_{\rm s}$. The corresponding mean free path is $\lambda_{\rm mfp} = 1/(n \sigma_{\rm t})$.

The integrals in Equations~\ref{eq:stop} and \ref{eq:sigt} require evaluating differential cross sections at arbitrary electron energies. For tabulated data, this procedure requires interpolating between tabulated distributions. We use the method of quantile functions described in Appendix~\ref{sec:algo}.

\subsubsection{The EEDL database}

We model electron-atom interactions using data from the Electron-Photon Intersection Cross Sections database, version 2023 (EPICS2023), which is freely available online.\footnote{\href{https://www-nds.iaea.org/epics/}{www-nds.iaea.org/epics/}} We use the Evaluated Electron Data Library (EEDL) \citep{Cullen2017}, one of the three datasets in the database. The EEDL dataset includes cross section and energy loss data for interactions between electrons and atoms with atomic numbers in the range $1$ to $100$. The data are given in the Evaluated Nuclear Data File (ENDF) format \citep{Trkov&Brown2018}, which we parse into a \texttt{json} file using a \textsc{Python} script. 

The warm temperatures $\gtrsim 10^3~{\rm K}$ and magnetic fields $\sim 10^{-6}~{\rm G}$ in KN ejecta (see Sec.~\ref{sec:toy_model}) are different than the cold, unmagnetized atoms for which the EEDL dataset was constructed. However, lacking an equivalent dataset more tailored for astrophysical applications, we argue that the EEDL data is a reasonable approximation for our application. 

First, the magnetic fields are too weak to significantly affect the atomic physics. At $10^{-6}~{\rm G}$, the typical Zeeman energy shift is $10^{-14}~{\rm eV}$. Second, the $\beta$-particles are much hotter than the ejecta, so we can neglect the thermal motion of the atoms and approximate the center-of-mass frame for each electron-atom interaction by the local fluid reference frame.

KN ejecta are also expected to be lightly ionized (charges $\lesssim +4$) by the $\beta$-particles, especially in the nebular phase \citep{Hotokezaka+2021, Pognan+2022, Brethauer+2026, Chiba+2026}, but the EEDL database only includes neutral targets. In the following sections, we discuss corrections to the neutral atom cross sections in the EEDL database to account for the ejecta ionization. 

\subsubsection{Electron impact ionization}

EEDL includes the EII cross section and secondary electron spectrum as a function of primary electron energy for each electron subshell $nlj$ of each element. The energy lost by the primary electron is the sum of the energy of the secondary electron and the subshell binding energy, $\Delta E = E_{\rm sec} + I$, where $I$ is the binding energy.

We adjust the neutral EII cross section for ionized targets using the empirical Lotz scaling \citep{Bethe1930, Lotz1967}. The EII cross section for subshell $nlj$ and ionization stage $q$ is related to the neutral cross section by
\begin{equation}
    \sigma_{{\rm EII}, q,nlj}(E) = \frac{w_{q,nlj}}{w_{0,nlj}} \left( \frac{I_{0, nlj}}{I_{q, nlj}} \right)^2 \sigma_{{\rm EII},0,nlj}\left( \frac{I_{0,nlj}}{I_{q,nlj}} E \right) \,,
    \label{eq:lotz}
\end{equation}
where $E$ is the initial energy of the primary and $w_{q, nlj}$ and $I_{q, nlj}$ are the occupation number and binding energy respectively for subshell $nlj$ and ionization stage $q$. The ratio of ionization potentials in the argument of the neutral cross section ensures that the cross sections are related at equivalent energies in threshold units. We approximate the subshell ionization potentials with energy eigenvalues, invoking Koopmans' theorem \citep{Koopmans1934}. We compute these eigenvalues using the Hartree-Fock method implemented in the \textsc{CATS} code from Los Alamos, an adaptation of the atomic structure codes developed by \citet{Cowan1981}.

\begin{figure*}
    \centering
    \includegraphics[width=0.667\linewidth]{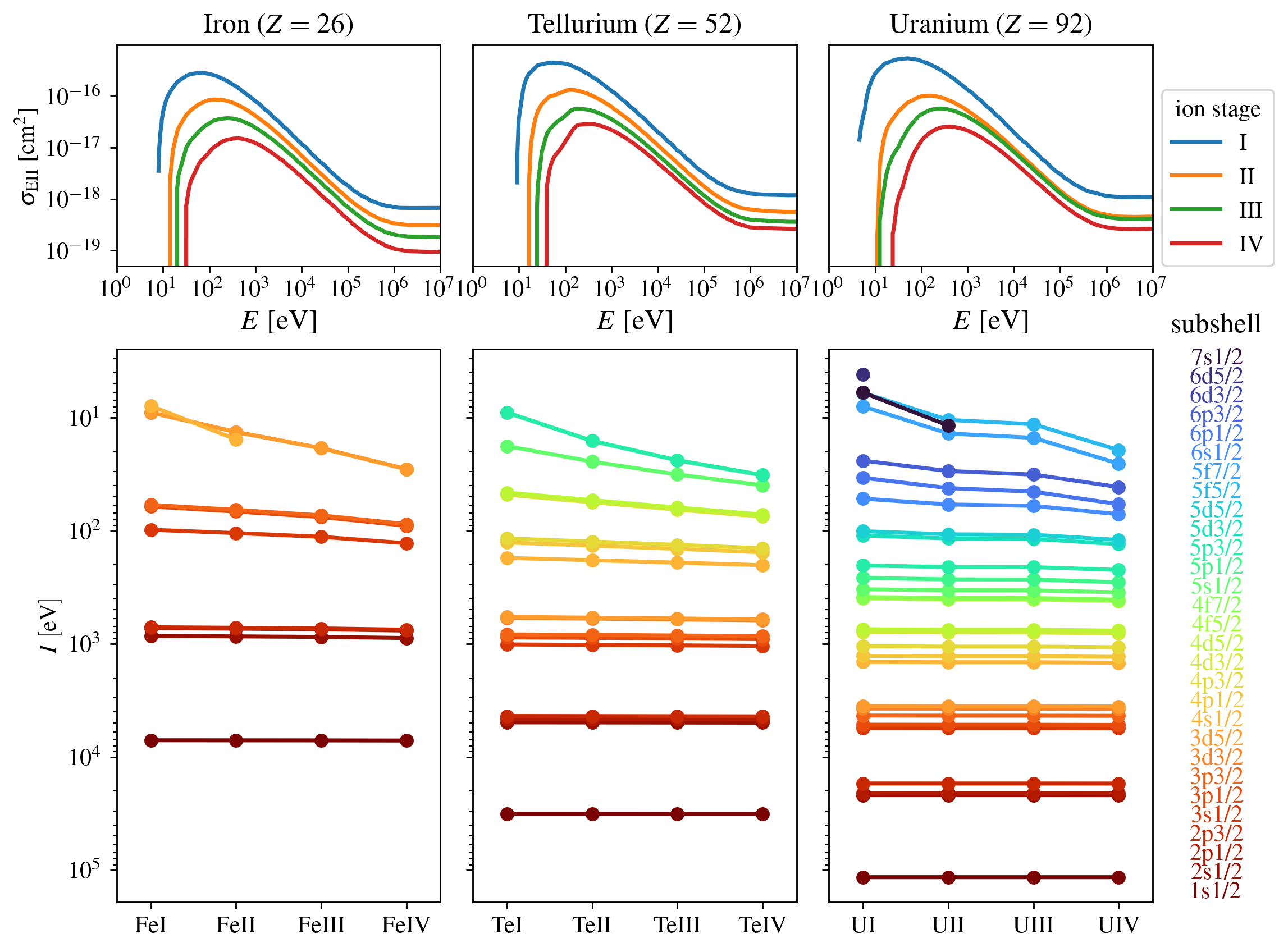}
    \caption{EII cross sections as a function of electron energy (top row) and subshell binding energies as a function of ionization stage (bottom row) for 3 elements with increasing atomic complexity: Iron ($Z=26$) (first column), Tellurium $(Z=52)$ (second column), and Uranium ($Z=92$) (third column). In the top row, each ionization stage is shown in a different color, indicated by the legend. In the bottom row, each subshell is shown in a different color indicated by the annotations on the right side of the plot. For an ion stage $q$, the EII cross section is smaller than its neutral value by a factor $\sim q^2$, although the precise energy-dependent adjustment to the cross section has a more complex dependence on the atomic structure.}
    \label{fig:sigion}
\end{figure*}

In Figure~\ref{fig:sigion}, we show the EII cross sections as a function of electron energy, and the subshell binding energies as a function of ionization stage $q$ for 3 elements with increasing atomic complexity: Iron ($Z=26$), Tellurium ($Z=52$), and Uranium ($Z=92$). The EII cross section at ion stage $q$ is smaller than its neutral value by a factor $\sim q^2$, although the precise energy-dependent adjustment has a more complex dependence on the atomic structure.

For a given subshell, the binding energy increases with ionization stage. We account for the extra energy by increasing the energy loss of the primary electron, while leaving the secondary electron distribution unchanged. This additive correction generally has a small effect on the EII stopping cross sections compared to the multiplicative correction from Equation~\ref{eq:lotz}.

\subsubsection{Electron impact excitation}
\label{sec:EIE}

EEDL includes the EIE cross section and energy loss as a function of electron energy for each element. In principle, the EIE cross sections should also be modified as a function of target ionization stage. However, EEDL does not partition the EIE cross section by subshell, so we cannot apply a scaling relation as we did for the EII cross sections. In Section~\ref{sec:total_xsec}, we show that the EIE stopping cross section contribution is subdominant relative to the EII contribution at energies $\gtrsim 1~{\rm keV}$, except for light elements like He where EIE dominates up to $\sim 100~{\rm keV}$. Therefore, neglecting the change in EIE cross sections has only a small effect on the stopping cross sections as $\beta$-particles thermalize from their initial MeV energies.

\subsubsection{Electron-atom bremsstrahlung}
\label{sec:brem}

The EEDL database includes the electron-atom bremsstrahlung cross section and average energy loss as a function of electron energy for each element. We apply these data to lightly ionized ejecta without modification. At impact parameters smaller than the atomic radius, this choice is justified because the missing electrons are spatially extended, so their absence does not significantly modify the Coulomb field probed by the $\beta$-particle. 

At impact parameters larger than the atomic radius, the contribution to the stopping cross section is subdominant. We can demonstrate this behavior heuristically by comparing analytic expressions for stopping cross section in the limits of no screening (ns) and full screening (fs) \citep{Bethe&Heitler1934, Koch&Motz1959}, i.e.
\begin{align}
    S_{\rm brem, ns} =\ & 4 r_0^2 \alpha q^2 \left[ \ln (2 \gamma) - \frac{1}{3} \right] \cdot E \label{eq:Sbrem_ns}\\
    S_{\rm brem, fs} =\ & 4 r_0^2 \alpha q^2 \left[ \ln\left(\frac{4}{3} \alpha^{-1} Z^{-1/3} \right) + \frac{1}{18} \right] \cdot E \label{eq:Sbrem_fs} \,,
\end{align}
where $r_0 \equiv e^2 / E_{\rm r}$ is the classical electron radius, $E_{\rm r} = m_{\rm e} c^2$ is the electron rest-mass energy, $\alpha \equiv e^2 / \hbar c$ is the fine-structure constant, $\gamma \equiv (1 - \beta^2)^{-1/2}$ is the particle Lorentz factor, $\beta \equiv v/c$, and these expressions are derived under the Born approximation in the ultra-relativistic limit. 

Equation~\ref{eq:Sbrem_ns} represents the stopping cross section for a bare nucleus with charge $q$, which is limited by the maximum impact parameter set by the photon formation length. Equation~\ref{eq:Sbrem_fs} represents the stopping cross section for a neutral atom with nuclear charge $q$, which is limited by the maximum impact parameter set by the atomic radius. For $E \lesssim 70\,Z^{-1/3}~{\rm MeV}$, which is satisfied for typical $\beta$-particle decay energies, the ns cross section is less than the fs cross section, so the photon formation length provides a more restrictive cutoff than the atomic radius and the large impact parameter contribution can be neglected.

\subsubsection{Mott scattering}
\label{sec:mott}

The EEDL database includes the differential Mott scattering cross section as a function of electron energy. By our arguments in the previous subsection, the transport cross section contribution at impact parameters smaller than the atomic radius is insensitive to target ionization.

For ionized targets, long-range Coulomb forces produce a small-angle scattering contribution that must be accounted for. In the limit of small $1 - \mu_{\rm s}$, where $\mu_s$ is the cosine of the scattering angle, the Mott scattering differential cross section for a target element $i$ is
\begin{equation}
    \dv{\sigma_{\rm mott}}{\mu_{\rm s}} \approx 2\pi \overline{q^2_i} \left( \frac{r_0}{\gamma \beta^2} \right)^2 \frac{1}{(1 - \mu_{\rm s})^2} \,,
    \label{eq:dsigdmu_mottsa}
\end{equation}
where $\overline{q^2_i}$ is the average square ionization stage of element $i$. The integrated cross section is
\begin{equation}
    \sigma_{\rm mott}(\mu_{\rm s}) = \int_{-1}^{\mu_{\rm s}} \dd \mu'_{\rm s} \dv{\sigma_{\rm mott}}{\mu'_{\rm s}} \approx 2\pi \overline{q^2_i} \left( \frac{r_0}{\gamma \beta^2} \right)^2 \frac{1}{1 - \mu_{\rm s}} \,.
\end{equation}
The scattering angle associated with impact parameter $b$ is given by setting the geometric area $\pi b^2$ equal to the integrated cross section $\sigma_{\rm mott}(\mu_{\rm s})$ and solving for $\mu_{\rm s}$, which gives
\begin{equation}
    1 - \mu_{\rm s} \approx \frac{2}{b^2} \overline{q^2_i} \left( \frac{r_0}{\gamma \beta^2}\right)^2 \,.
    \label{eq:b_to_mu_mott}
\end{equation}

Small-angle scattering is produced by impact parameters from the atomic radius to the Debye length, given by
\begin{equation}
    \lambda_{\rm D} = \left(\frac{k_{\rm B} T}{4\pi e^2 n_{\rm i} (\overline{q} + \overline{q^2})}\right)^{1/2} \simeq 2.2~T_3^{1/2} n_{{\rm i}, 4}^{-1/2} (\overline{q} + \overline{q^2})^{-1/2}~{\rm cm} \,,
    \label{eq:debye}
\end{equation}
where $T_3 = T/10^3~{\rm K}$ is the temperature, $n_{{\rm i}, 4} = n_{\rm i} / 10^4~{\rm cm^{-3}}$ is the ion number density, and $\overline{q}$ and $\overline{q^2}$ are the average ionization stage and square ionization stage across all elements. We have included the contribution of free electrons, whose number density is $\overline{q} n_{\rm i}$. On scales $\gtrsim \lambda_{\rm D}$, the Coulomb potential of the ion is effectively screened by polarization of the ejecta and the associated redistribution of charge.

Computing the transport cross section (Eq.~\ref{eq:sigt}) for the Mott differential cross section (Eq.~\ref{eq:dsigdmu_mottsa}) and integrating over the allowed impact parameters gives
\begin{equation}\begin{split}
    \sigma_{\rm t, Mott} =\ & 4\pi \overline{q^2_i} \left( \frac{r_0}{\gamma \beta^2} \right)^2 \ln \Lambda_{\rm sa}\\
    \ & \simeq 1.0\times 10^{-24}~\overline{q^2_i} \gamma^{-2} \beta^{-4} \ln\Lambda_{{\rm sa}}~{\rm cm^2} \,,
\end{split}\end{equation}
where $\ln \Lambda_{\rm sa} = \ln(\lambda_{\rm D}/r_{\rm atom}) \simeq \ln \Lambda - 6.8 + (1/3)\ln Z$ is the small-angle Coulomb logarithm, which we define in relation to the total Coulomb logarithm
\begin{equation}
	\ln \Lambda = \ln (\frac{\lambda_{\rm D}}{\lambda_{\rm C}}) \simeq 22.9 + \frac{1}{2} \ln(\frac{T_3}{n_{\rm i, 4} (\overline{q} + \overline{q^2})}) \,,
	\label{eq:coulog}
\end{equation}
where $\lambda_{\rm C} = h/m_{\rm e}c$ is the electron Compton wavelength.

\subsection{M{\o}ller scattering}
\label{sec:moller}

In addition to Coulomb interactions with atoms and ions, $\beta$-particles also lose energy through Coulomb interactions with free thermal electrons. The differential cross section for electron-electron scattering, or M{\o}ller scattering, has an analytic form from quantum field theory given by \citep[][Eq.~6-42]{Moller1932, Itzykson&Zuber1980}
\begin{equation}\begin{split}
    \dv{\sigma_{\rm mol}}{\Omega_{\rm cm}} =\ & \frac{e^4 ( 2 \mathcal{E}_{\rm cm}^2 - E_{\rm r}^2)^2}{4 \mathcal{E}_{\rm cm}^2 (\mathcal{E}_{\rm cm}^2 - E_{\rm r}^2)^2}\\
    \ & \left[ \frac{4}{\sin^4 \theta_{\rm cm}} - \frac{3}{\sin^2 \theta_{\rm cm}} + \frac{(\mathcal{E}_{\rm cm}^2 - E_{\rm r}^2)^2}{(2 \mathcal{E}_{\rm cm}^2 - E_{\rm r}^2)^2} \left( 1 + \frac{4}{\sin^2 \theta_{\rm cm}}\right) \right] \,,
    \label{eq:moller1}
\end{split}\end{equation}
where $\mathcal{E} = E + E_{\rm r}$ is the total electron energy, $E$ is the incident electron energy, $E_{\rm r} = m_{\rm e} c^2$ is the electron rest-mass energy, and the subscript ${\rm ``cm"}$ indicates that a quantity is measured in the center-of-mass (CM) frame. The Lorentz factors in the CM and local fluid frames are related by \citep[][Eq.~67]{Moller1932},
\begin{equation}
    \gamma_{\rm cm} = \sqrt{(1/2)(\gamma + 1)} \,.
    \label{eq:gammacm}
\end{equation}
Using Equation~\ref{eq:gammacm} to write $\mathcal{E}_{\rm cm}$ in terms of the fluid frame Lorentz factor, we recast Equation~\ref{eq:moller1} in the following form,
\begin{equation}\begin{split}
    \dv{\sigma_{\rm mol}}{\mu_{\rm s,cm}} =\ & 4\pi \left( \frac{r_0}{\gamma \beta} \right)^2 \frac{\gamma + 1}{\beta^2}\\
    \ & \left[ \frac{4}{(1 - \mu_{\rm s,cm}^2)^2} - \frac{3}{1 - \mu_{\rm s,cm}^2} + \frac{(\gamma - 1)^2}{4 \gamma^2} \left( 1 + \frac{4}{1 - \mu_{\rm s,cm}^2} \right) \right] \,.
    \label{eq:moller2}
\end{split}\end{equation}
The scattering angles in the CM and fluid frames are related by a Lorentz boost via \citep[][Eq.~68]{Moller1932}
\begin{equation}
    \mu_{\rm s,cm} = \frac{\mu_{\rm s}^2 - \gamma_{\rm cm}^2 (1 - \mu_{\rm s}^2)}{\mu_{\rm s}^2 + \gamma_{\rm cm}^2 (1 - \mu_{\rm s}^2)} = \frac{2 - (\gamma + 3) (1 - \mu_{\rm s}^2)}{2 + (\gamma - 1) (1 - \mu_{\rm s}^2)} \,.
\end{equation}
The inverse relation is given by
\begin{equation}
    \mu_{\rm s}^2 = \frac{(1 + \gamma)(1 + \mu_{\rm s, cm})}{(1 + \gamma)(1 + \mu_{\rm s, cm}) + 2 (1 - \mu_{\rm s, cm}) }
    \label{eq:mu_cm_to_mu} \,.
\end{equation}

In the CM frame, the two electrons produced by a M{\o}ller scattering interaction are identical, producing the forward-backward symmetry in the differential cross section. In other words, an electron is equally likely to scatter forwards or backwards in the CM frame. In the fluid frame, the forward-scattered electron has more energy than the backward-scattered electron. To ensure that we always follow the higher-energy electron, we multiply the differential cross section by two and restrict its domain to $\mu_{\rm s,cm} > 0$. It follows that the energy loss in the local fluid frame is at most half the electron energy. Doing a Lorentz boost, the precise formula is
\begin{equation}
    \Delta E / E = (1/2) (1 - \mu_{\rm s,cm})
    \label{eq:moller_loss} \,.
\end{equation}
This is different from Mott scattering, where the CM and fluid frames nearly coincide because the target nucleus has mass $\gg m_{\rm e}$. Consequently, if scattering is elastic in the CM frame, there is only a small energy exchange even for a significant deflection of the incident electron.

The minimum impact parameter is the relativistic de Broglie wavelength of the $\beta$-particle given by $\lambda_{\rm e} = h c / (\gamma E_{\rm r} \beta)$. On scales $\lesssim \lambda_{\rm e}$, $\beta$-particles cannot be localized due to the Heisenberg uncertainty principle. The maximum impact parameter is the Debye length of the ejecta (Eq.~\ref{eq:debye}).  

To leading order in $1 - \mu_{\rm cm}$, the integrated cross section is
\begin{equation}\begin{split}
    \sigma_{\rm mol}(\mu_{\rm s,cm}) =\ & 2 \int_0^{\mu_{\rm s,cm}} \dd \mu_{\rm cm}' \dv{\sigma_{\rm mol}}{\mu_{\rm s,cm}'}\\
    \approx\ & 8\pi \left( \frac{r_0}{\gamma \beta} \right)^2 \frac{\gamma + 1}{\beta^2} \frac{1}{1 - \mu_{\rm s,cm}} \,.
    \label{eq:sig_moller}
\end{split}\end{equation} 
Similar to our calculation of Mott scattering (Sec~\ref{sec:mott}), the scattering angle and impact parameter are related by $1 - \mu_{\rm s, cm} \propto b^{-2}$. Computing the stopping cross section (Eq.~\ref{eq:stop}) for the M{\o}ller differential cross section (Eq.~\ref{eq:moller2}) and energy loss (Eq.~\ref{eq:moller_loss}) to leading order in $1 - \mu_{\rm s,cm}$ gives
\begin{equation}\begin{split}
    S_{\rm mol} \approx\ & 8\pi \left( \frac{r_0}{\gamma \beta} \right)^2 \frac{\gamma + 1}{\beta^2} \ln \Lambda' \cdot E\\
    \simeq\ & 2.0\times 10^{-24}~(\gamma + 1) \gamma^{-2} \beta^{-4}\ln \Lambda'~{\rm cm^2} \cdot E \,,
    \label{eq:moller_stop}
\end{split}\end{equation}
where $\ln \Lambda' = \ln (\lambda_{\rm D} / \lambda_{\rm e}) = \ln \Lambda +  \ln(\gamma \beta)$ and $\ln \Lambda$ is given by Equation~\ref{eq:coulog}. Computing the transport cross section (Eq.~\ref{eq:sigt}) for the M{\o}ller differential cross section and fluid frame scattering angle (Eq.~\ref{eq:mu_cm_to_mu}) to leading order in $1 - \mu_{\rm s,cm}$ gives
\begin{equation}\begin{split}
    \sigma_{\rm t,mol} \approx\ & 8\pi \left( \frac{r_0}{\gamma \beta^2} \right)^2 \ln \Lambda'\\
    \simeq\ & 2.0\times 10^{-24}~\gamma^{-2} \beta^{-4}\ln \Lambda'~{\rm cm^2} \,,
    \label{eq:moller_stop}
\end{split}\end{equation}
which only differs from the stopping cross section by a factor $(\gamma + 1) E$.

\subsection{Total stopping and transport cross sections}
\label{sec:total_xsec}

\begin{figure*}
    \centering
    \includegraphics[width=\linewidth]{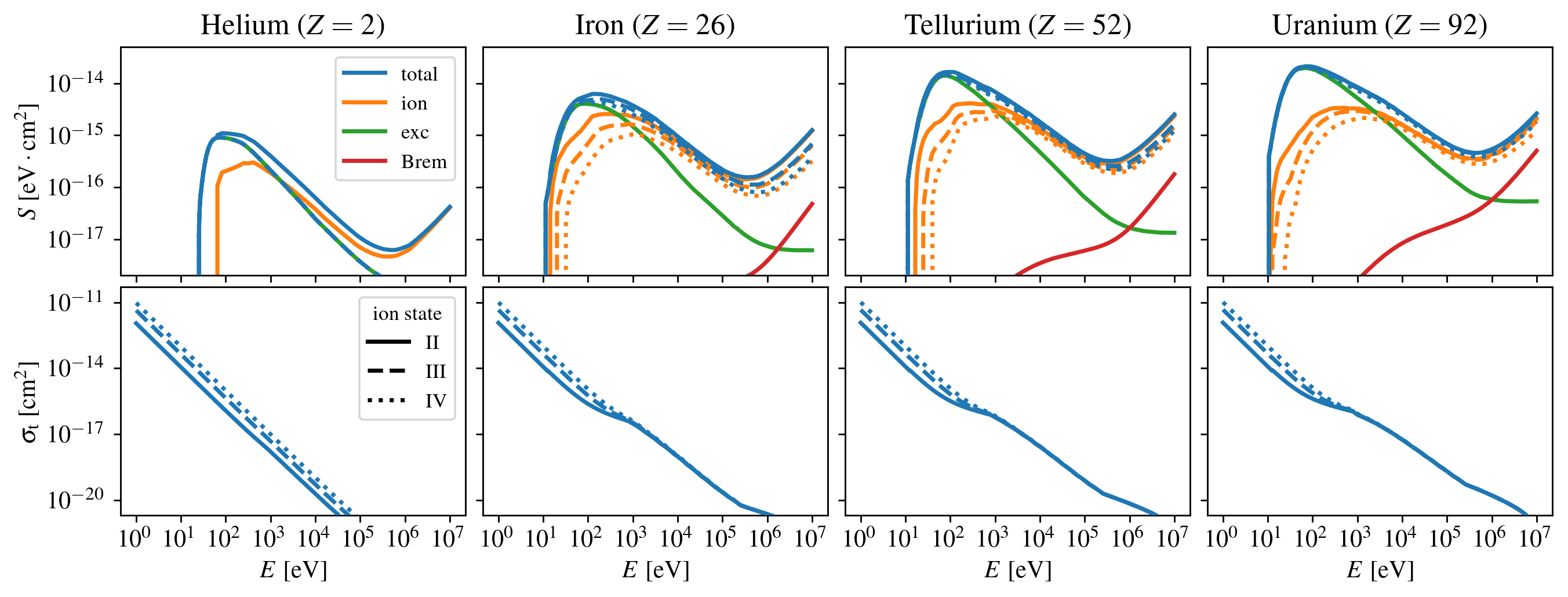}
    \caption{Stopping cross section (top row) and transport cross section (bottom row) for Helium ($Z=2$) (first column), Iron ($Z=26$) (second column), Tellurium ($Z=52$) (third column), and Uranium ($Z=92$) (fourth column) as a function of electron energy. The stopping cross section is partitioned into contributions from EII (orange), EIE (green), and bremsstrahlung (red). Each target ionization stage is shown in a different linestyle indicated by the legend. Energy loss is driven by EII at MeV energies, but EIE dominates at lower energies $\lesssim 1~{\rm keV}$. Bremsstrahlung does not contribute significantly to the stopping cross section at MeV energies, although it dominates at GeV energies beyond the limit of the plot. At low energies $\lesssim 1~{\rm keV}$, the transport cross section is dominated by the small-angle Mott contribution, so it scales as $E^{-2}$ and is sensitive to the target ionization stage. At higher energies, the transport cross section is augmented by the large-angle Mott contribution, so it is insensitive to the target ionization stage. Heavier species have higher transport cross sections, higher stopping cross sections, and greater relative contributions from EIE and bremsstrahlung. }
    \label{fig:EEDL}
\end{figure*}

In Figure~\ref{fig:EEDL}, we show the stopping and transport cross sections for Helium ($Z=2$), Iron ($Z=26$), Tellurium ($Z=52$), and Uranium ($Z=92$) targets as a function of electron energy and ionization stage. The stopping cross section is partitioned into contributions from EII, EIE, and bremsstrahlung.

Energy loss is driven by EII at MeV energies, but EIE dominates at lower energies $\lesssim 1~{\rm keV}$. At the first ionization threshold, the EII contribution rises sharply, while the EIE contribution is negligible until somewhat higher energies where strong dipole-allowed excitation channels become available. Bremsstrahlung does not contribute significantly to the stopping cross section at MeV energies, although it dominates at GeV energies beyond the limit of the plot. Heavier species have higher stopping cross sections and greater relative contributions from EIE and bremsstrahlung. 

At low energies $\lesssim 1~{\rm keV}$, the transport cross section is dominated by the small-angle Mott contribution, so it scales as $E^{-2}$ and is sensitive to the target ionization stage. At higher energies, the transport cross section is augmented by the large-angle Mott contribution, so it is insensitive to the target ionization stage and increases for heavier species.

If we let $x_{iq}$ denote the number fraction of element $i$ in ionization stage $q$, then the total stopping and transport cross sections can be expressed as an abundance-weighted sum over the contributions from each species according to
\begin{align}
	S_{\rm tot} =\ & \mu_{\rm mmw} \sum_i Y_i \sum_q x_{iq} S_{iq} + \overline{q} S_{\rm e} \label{eq:Stot}\\
    \sigma_{\rm t, tot} =\ & \mu_{\rm mmw}\sum_i Y_i \sum_q x_{iq} \sigma_{{\rm t}, iq} + \overline{q} \sigma_{\rm t,e} \label{eq:sigttot} \,,
\end{align}
where $S_{i q}$ and $\sigma_{{\rm t}, iq}$ are the stopping and transport cross sections for species $(i, q)$, and $S_{\rm e}$ and $\sigma_{\rm t, e}$ are the electron stopping and transport cross sections due to M{\o}ller scattering. These quantities are related to the mass stopping power and transport opacity by
\begin{equation}
    -\frac{1}{\rho} \dv{E}{x} = \frac{N_{\rm A}}{\mu_{\rm mmw}} S_{\rm tot}, \quad \kappa_{\rm t} = \frac{N_{\rm A}}{\mu_{\rm mmw}} \sigma_{\rm t, tot} \,.
\end{equation}

\begin{figure}
    \centering
    \includegraphics[width=\linewidth]{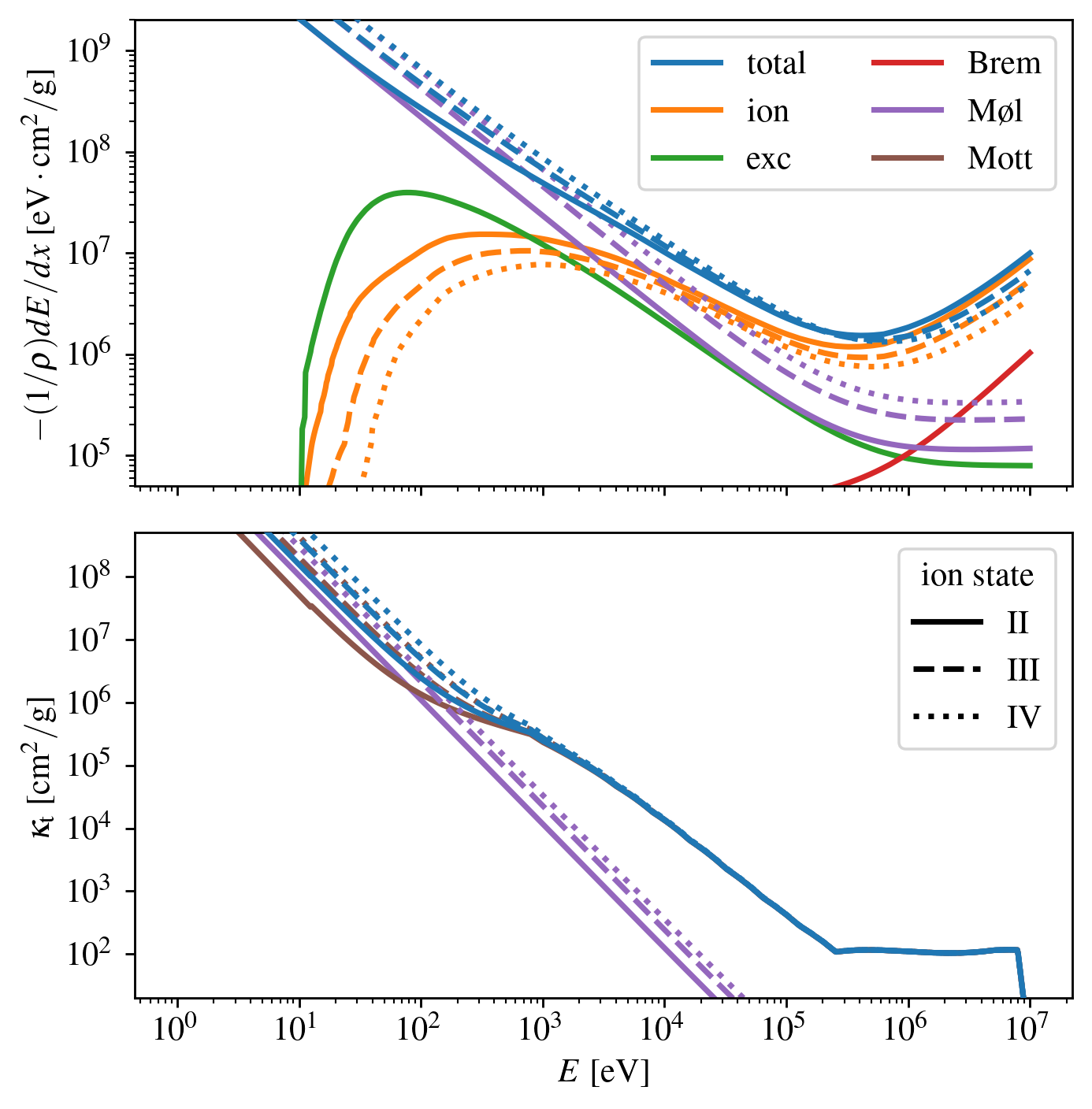}
    \caption{Mass stopping power (top panel) and transport opacity (bottom row) as a function of electron energy for our strong $r$-process abundance at 30~{\rm days} post-merger. The mass stopping power is partitioned into contributions from EII (orange), EIE (green), bremsstrahlung (red), and M{\o}ller scattering (purple). The transport opacity is partitioned into contributions from Mott scattering (brown) and M{\o}ller scattering (purple). Each target ionization stage is shown in a different linestyle indicated by the legend. The results are similar to the results for individual elements (Fig.~\ref{fig:EEDL}), except for the contributions from M{\o}ller scattering. These contributions dominate at low energies $\lesssim 1~{\rm keV}$ and become stronger at higher ionization stages. Low energy $\beta$-particles rapidly thermalize due to Coulomb interactions with thermal electrons.}
    \label{fig:stop_kap}
\end{figure}

In Figure~\ref{fig:stop_kap}, we plot the total mass stopping power and transport opacity as a function of electron energy for our strong $r$-process abundance at 30~{\rm days} post-merger. The results are similar to the results for individual elements (Fig.~\ref{fig:EEDL}), except for the contributions from M{\o}ller scattering. These contributions dominate at low energies $\lesssim 1~{\rm keV}$ and become stronger at higher ionization stages, suggesting that low energy $\beta$-particles rapidly thermalize due to Coulomb interactions with thermal electrons.

There are several microphysical processes not included in our modeling: Auger-Meitner ionization, Compton scattering, synchrotron losses, and Cherenkov radiation. We discuss these in Appendix~\ref{sec:not_modeled}, and argue that including them would not significantly modify our results. The exception is Cherenkov radiation losses (App.~\ref{sec:cher}), for which the mass stopping power is $\sim 10\%$ of the total mass stopping power at energies $\lesssim 10~{\rm keV}$, and which we do not include due to modeling uncertainties.

\subsection{Ionization cascade}
\label{sec:cascade}

EII generates a population of secondary electrons with a wide energy distribution that peaks at $\sim100~{\rm eV}$ and has a long tail at higher energies. As the secondary population thermalizes, it produces a population of tertiary electrons. Iterating this process results in an ionization cascade. The typical secondary electron produces far fewer ionizations than the typical primary. However, there are many secondary electrons for each primary electron and secondary electrons are produced closer to threshold energies where the EII cross sections peak, so the ionization cascade can contribute significantly to the overall ionization rate.

As an electron of energy $E$ thermalizes, the number of times it ionizes a species $(i, q)$ is
\begin{equation}
    N_{{\rm ion}, iq}(E) = \int \dd E'  \frac{\Gamma_{{\rm ion}, iq}}{\abs{\dd E/\dd t}} = Y_{i} x_{iq} \mu_{\rm mmw} \int_{0}^{E} \dd E' \frac{\sigma_{{\rm EII}, iq}(E')}{S_{\rm tot}(E')} \,,
    \label{eq:nion}
\end{equation}
where $\Gamma_{{\rm ion}, iq} = n_{iq} v \sigma_{{\rm EII}, iq}$ is the ionization rate of species $(i, q)$. Similarly, as an electron at energy $E$ thermalizes, the spectrum of secondaries it produces at energy $E'$ is
\begin{equation}
    \eval{\dv{N_{\rm sec}(E)}{E'}}_{E'} = \int_{0}^E \frac{\dd E''}{S_{\rm tot}(E'')} \dv{\sigma_{\rm sec}(E'')}{E'} \,,
    \label{eq:nsec}
\end{equation}
where $\dd \sigma_{\rm sec}(E) / \dd E'$ is the differential cross section for an electron at energy $E$ to produce a secondary electron at energy $E'$, including contributions from all target species.

Convolving the number of ionizations with the differential cross section for secondaries, we get the effective ionization cross section due to secondaries,
\begin{equation}
    \sigma_{{\rm ion}, iq}(E) = \int \dd E' N_{{\rm ion}, iq}(E') \frac{\dd \sigma_{\rm sec}(E)}{\dd E'} \,.
    \label{eq:sig_ion_sec}
\end{equation}
By using this modified cross section, we assume that secondary electrons thermalize locally in space and time, similar to assumptions in previous work \citep{Barnes+2016, Shenhar+2024}. Under this assumption, the ionization cascade only affects the local ionization rate, while the EII stopping power remains unchanged.

Convolving the secondary spectrum with the differential cross section for secondaries, we get the differential cross section for tertiaries,
\begin{equation}
    \dv{\sigma_{\rm ter}(E)}{E'} = \int \dd E'' \dv{N_{\rm sec}(E'')}{E'} \dv{\sigma_{\rm sec}(E)}{E''} \,.
    \label{eq:sig_ter}
\end{equation}

\begin{figure}
    \centering
    \includegraphics[width=1.0\linewidth]{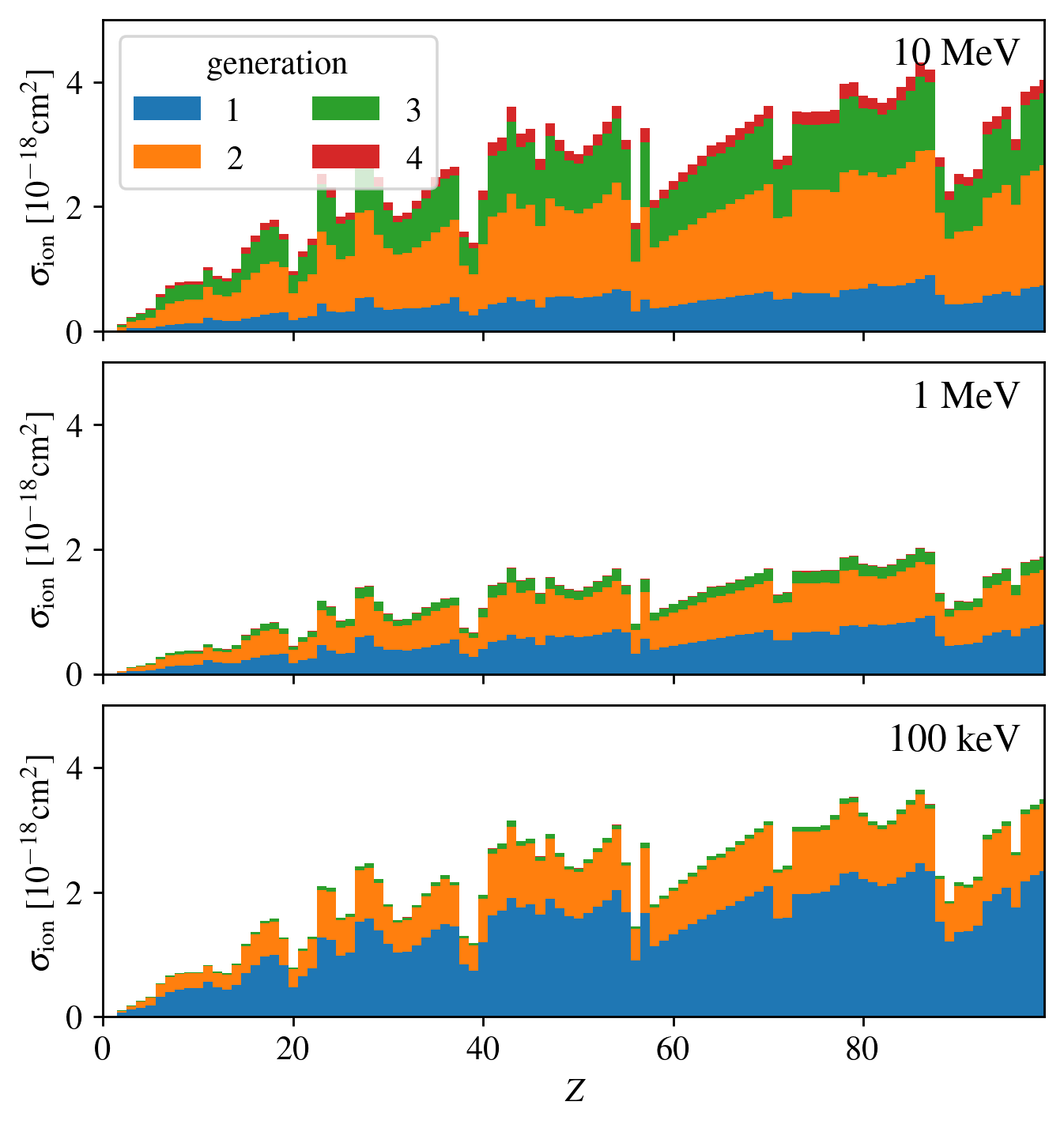}
    \caption{Per-element effective electron impact ionization (EII) cross sections for singly-ionized targets and three incident electron energies: 10~MeV (top panel), 1~MeV (middle panel), 100~keV (bottom panel). Cross sections are partitioned into contributions from each generation of the ionization cascade (Sec.~\ref{sec:cascade}): primaries (blue), secondaries (orange), tertiaries (green), and quaternaries (red). The cascade (secondaries and beyond) contributes significantly to the EII cross section across all elements. The cascade contribution increases with incident electron energy, becoming similar to the primary contribution at $1~{\rm MeV}$.}
    \label{fig:sigsec}
\end{figure}

These calculations can be iterated for successive generations by substituting the result of Equation~\ref{eq:sig_ter} for $\dd \sigma_{\rm sec}(E) / \dd E'$ in Equations~\ref{eq:sig_ion_sec} and \ref{eq:sig_ter}. The energy in the cascade (secondaries and beyond) decreases with each successive generation due to the energy spent to overcome atomic binding energies and non-EII energy loss channels like EIE and M{\o}ller scattering. In Figure~\ref{fig:sigsec}, we show the effective EII cross section across singly-ionized targets partitioned into contributions from each generation of the ionization cascade. We find that the cascade contributions are significant across all elements and increase with incident electron energy, becoming comparable to the primary contribution at $\sim 1~{\rm MeV}$. In a recent work, \citet{VandenBerg&Hotokezaka2026} also conclude that ionization rates are significantly enhanced by the cascade using a Spencer-Fano framework.

The enhancement to the effective EII cross section can be approximately described by an empirical fit
\begin{equation}
	\sigma_{\rm ion, cas} \simeq \sigma_{\rm ion} \left[ 1 + A \left( \frac{E}{{\rm MeV}}\right)^{1/2} \right] \,,
	\label{eq:cascade_formula}
\end{equation}
where $\sigma_{\rm ion, cas}$ and $\sigma_{\rm ion}$ are the effective EII cross sections with and without the cascade contribution respectively and $A \simeq 1.5$ is a constant. The enhancement is typically greater for lower ionization targets because their loosely bound valence electrons are more easily ionized by cascade electrons. Our approximate formula is within ~30\% of the true enhancement $\sigma_{\rm ion, cas}/\sigma_{\rm ion}$ across all target species and incident electron energies. For a more accurate description of the enhancement, our public data include the ionization cross sections with and without the cascade contribution for all target species (see the \hyperref[sec:data]{Data Availability statement}).

We also expect cascade electrons to enhance the effective excitation rate. However, given our simplified treatment of EIE, we leave a more detailed analysis of this effect to future work that models the ejecta excitation state more self-consistently. We note that by neglecting the effect of target ionization on the EIE stopping power (Sec.~\ref{sec:EIE}), we underestimate the number of ionizations over the lifetime of a cascade electron. However, we expect this effect to be small because the stopping power contribution from M{\o}ller scattering always dominates over that from EIE (Fig.~\ref{fig:stop_kap}).

\section{Global model}
\label{sec:global_model}
\subsection{Toy model}
\label{sec:toy_model}

The dynamical ejecta are expected to have masses in the range $10^{-4}-10^{-2}~M_\odot$, velocities in the range $0.1-0.3 c$, and electron fractions in the range $0.05-0.4$ based on theoretical work \citep{Rosswog+1999, Rosswog2013, Bauswein+2013, Hotokezaka+2013, Wanajo+2014, Kyutoku+2015, Sekiguchi+2016, Radice+2018}. The mass and velocity ranges are broadly consistent with modeling of AT2017gfo \citep{Smartt+2017, Kasliwal+2017b, Abbott+2017a, Cowperthwaite+2017, Rosswog+2018}.

After the merger, the ejecta expands rapidly. As the flow becomes supersonic, the expansion becomes approximately homologous. This behavior arises naturally in simulations \citep{Rosswog+2014, Grossman+2014} and is often assumed in light-curve calculations \citep{Metzger+2010, Kasen+2017, Tanaka+2017, Bulla2019, Kawaguchi+2020, Korobkin+2021}, yielding similar results to full radiation-hydrodynamic calculations \citep{Wu+2022}.

In the homologous model, at time $t$ post-merger, the ejecta extends out to a radius
\begin{equation}
    R_{\rm ej} = v_{{\rm ej}} t \simeq 5.2\times 10^{14}~v_{0.2} t_{\rm d}~{\rm cm} \,,
    \label{eq:Rej_toy}
\end{equation}
where $v_{0.2} \equiv v_{\rm ej} / 0.2c$, $t_{\rm d}$ is the time in days, and $v_{\rm ej}$ is the maximum ejecta velocity. The average density of the ejecta is
\begin{equation}
    \overline{\rho} = (3/4) \pi^{-1} M_{\rm ej} R_{\rm ej}^{-3} \simeq 3.4 \times 10^{-14}~M_{0.01} v_{0.2}^{-3} t_{\rm d}^{-3}~{\rm g/cm^3} \,,
    \label{eq:rho_toy}
\end{equation}
where $M_{0.01} \equiv M_{\rm ej}/10^{-2}M_\odot$.

We assume that the ejecta mass is distributed in velocity space according to a power law given by $\dd M / \dd u \propto u^{2-\alpha}$, where $u$ is the local ejecta velocity and $\alpha \sim 0.5 - 3.0$ \citep{Bauswein+2013, Metzger2019, Fryer+2024}. Under homologous expansion, $u \propto r$, so this approach is equivalent to a radial density profile $\rho \propto r^{-\alpha}$. 

Modeling of AT 2017gfo suggests that the effective temperature scales as $t^{-\beta}$, where $\beta \sim 0.5 - 0.8$, reaching $300 - 10^4~{\rm K}$ at late times \citep{Drout+2017, Waxman+2018, Sneppen+2024, Fryer+2024}. However, radiative transfer simulations show the gas temperature does not follow this scaling, and instead increases at late times \citep{Pognan+2022, Brethauer+2026}. The temperature only affects the microphysics weakly through the Coulomb logarithm (Sec.~\ref{sec:microphysics}), so we adopt a nominal value of $10^3~{\rm K}$.

If the magnetic flux is frozen into the ejecta, then the field strength is \citep{Barnes+2016}
\begin{equation}
    \overline{B} = B_0 (R_0 / R_{\rm ej})^2 \simeq 3.7 ~B_{12} R_6^2 v_{0.2}^{-2} t_{\rm d}^{-2}~{\rm \mu G} \,,
    \label{eq:B_toy}
\end{equation}
where $B_0$ and $R_0$ are the magnetic field strength and radius at the time of mass ejection, $B_{12} \equiv B_0 / 10^{12}~{\rm G}$ and $R_6 \equiv R_0 / 10^6~{\rm cm}$. $B_0$ ranges from $10^9$ to $10^{15}~{\rm G}$ depending on the initial NS magnetic fields and the efficiency of magnetic field amplification \citep{Price&Rosswog2006, Kiuchi+2015, Kiuchi+2018}. $R_0$ can be interpreted as the characteristic size of the neutron stars or the post-merger accretion disk, which ranges from $10^6$ to $10^7~{\rm cm}$.

In a magnetic field, the motion of charged particles is affected by the Lorentz force. For a constant or slowly-varying field, particle trajectories are described by circular motion about a guiding center that moves parallel to the field. These gyro-oscillations have radius $r_{\rm g} = \gamma m_{\rm e} v c \sqrt{1 - \mu^2} / (e B)$, where $\mu$ is the cosine pitch angle between the velocity and magnetic field. In the non-relativistic (NR) and ultra-relativistic (UR) limits, the gyro-to-ejecta radius ratio is
\begin{equation}
	r_{\rm g} / R_{\rm ej} \simeq 1.7 \times 10^{-6} B_{12}^{-1} R_6^{-2} v_{0.2} t_{\rm d} \begin{cases} E_{\rm MeV}^{1/2} & {\rm NR}\\ E_{\rm MeV} & {\rm UR}  \,,\end{cases}
\end{equation}
where $E_{\rm MeV} = E / {\rm MeV}$. For most realistic parameters, $\beta$-particles are well-magnetized and are transported parallel to the magnetic field. This behavior makes the thermalization efficiency sensitive to the magnetic field geometry \citep{Barnes+2016}. 

If magnetic fields are radial, then charged particles stream out of the ejecta and deposit less energy. If magnetic fields are toroidal or highly turbulent, then charged particles may be trapped in the ejecta. These effects have been previously discussed in the context of positron emission from core collapse supernovae \citep{Colgate+1980, Chan&Lingenfelter1993, Milne+1999}, motivated by the need to explain the strength of the Galactic gamma-ray emission line at $511~{\rm keV}$ produced by positron annihilation \citep{Siegert2023}. In these systems, radio \citep{Dickel&Milne1976, Milne1987, Dubner&Giacani2015} and X-ray polarimetry \citep{Vink+2022} favor radial magnetic fields in young supernova remnants, with an increasing turbulent component as the remnants age. While the physical conditions in KNe differ, these observations motivate considering a variety of magnetic field morphologies in the ejecta.

\subsection{Transport equations}
\label{sec:transport_equations}

We use Monte Carlo simulations to model $\beta$-particle transport and thermalization in the ejecta. We model the ejecta as a spherically symmetric, homologously expanding flow, which reduces the problem to one spatial dimension in radius. It is convenient to work in velocity coordinates $u = r/t$, which are constant along a Lagrangian position in the ejecta. We define two reference frames: the lab frame, the rest frame of the ejecta center-of-mass; and the local comoving frame (LCF), the rest frame of the fluid at the location of a $\beta$-particle. We denote LCF quantities with primes and lab-frame quantities without primes.

Electron transport in tangled magnetic fields is a complex process involving guiding center drifts, wave-particle interactions, and resonant scattering \citep[for a review, see][]{Verscharen+2026}. Given the uncertainty in the magnetic field structure, we consider two limiting cases for transport rather than modeling electron transport in detail: particles moving through coherent radial magnetic fields and particles that are completely trapped. 

Radial fields maximize particle transport because particles stream along radial field lines and are accelerated by magnetic focusing. For this scenario, we show that the 1D transport equations are mathematically equivalent to a scenario with no magnetic field. 

The completely trapped limit could arise in toroidal magnetic fields that trap particles at a single velocity coordinate, or in highly tangled magnetic fields that trap particles in small-scale magnetic structures. For more realistic field geometries, like the ``random field'' model considered by \citet{Barnes+2016}, particle transport is expected to produce results in-between these two extremes. Magnetized turbulence may also modify the electron spectrum through particle acceleration processes such as diffusive shock acceleration, but these effects are beyond the scope of this work.

It is convenient to work with dimensionless components of the particle 4-momentum: the particle Lorentz factor $\gamma$ and the dimensionless momenta $p_\parallel = \gamma v \mu$ and $p_\perp = \gamma v (1-\mu^2)^{1/2}$, where $v$ is the particle velocity in units of $c$ and $\mu$ is the cosine angle between the velocity and the radial direction. They are related by
\begin{equation}
	\gamma^2 = 1 + p_\parallel^2 + p_\perp^2
	\label{eq:einstein}
\end{equation}
and their values between reference frames are related by a Lorentz transformation, i.e.
\begin{align}
	\gamma =\ & \gamma_u (\gamma' + u p_\parallel') \quad \gamma' = \gamma_u (\gamma - u p_\parallel) \label{eq:gam_lt}\\
	p_\parallel =\ & \gamma_u (p_\parallel' + u \gamma') \quad p_\parallel' = \gamma_u (p_\parallel - u \gamma) \label{eq:ppar_lt}\\
	p_\perp =\ & p_\perp' \,,\label{eq:pperp_lt}
\end{align}
where $u = r/t$ is the fluid velocity in units of $c$ and $\gamma_u = (1-u^2)^{-1/2}$ is the fluid Lorentz factor. Time between reference frames is related by the Doppler factor given by
\begin{equation}
	D \equiv \dv{t}{t'} = \frac{\gamma}{\gamma'} = \gamma_u \frac{\gamma' + u p_\parallel'}{\gamma'} \,.
\end{equation}

As the particle travels, it experiences a velocity gradient given by
\begin{equation}
	\dv{u}{t'} = D\frac{\mu v - u}{t} = D\frac{p_\parallel - u \gamma}{\gamma t} = \frac{p_\parallel'}{\gamma_u \gamma' t} \,.
	\label{eq:udot}
\end{equation}
As a result, the energy and momentum in the LCF evolve, even in the absence of collisional losses. In the absence of a magnetic field, particles stream freely and the parallel momentum evolves due to the rotation of the radial unit vector along straight-line trajectories
\begin{equation}
	\eval{\dv{p_\parallel}{t'}}_{\rm geo} = D \dv{\vu{r}}{t} \vdot \vb{p} = D \frac{p_\perp^2}{\gamma u t} = \frac{p_\perp'^2}{\gamma' u t} \,.
	\label{eq:ppardot_ad}
\end{equation}
In a coherent radial magnetic field frozen into the ejecta, flux conservation implies $B \propto r^{-2}$. Assuming that the electrons are well-magnetized, the resulting field gradient produces a lab-frame mirror force \citep{Kulsrud2005}
\begin{equation}
	\eval{\dv{p_\parallel}{t'}}_{\rm mir} = -D \mu_{\rm mag} \dv{B}{r} = D \frac{p_\perp^2}{\gamma u t} = \frac{p_\perp'^2}{\gamma' u t} \,,
\end{equation}
where $\mu_{\rm mag} = p_\perp^2 / (2 \gamma B)$ is the magnetic moment. This result is the same as Equation~\ref{eq:ppardot_ad}, so this scenario is mathematically equivalent to the free-streaming case. In this configuration, the magnetic field and the fluid velocity are parallel, so there is no motional electric field and thus no lab-frame energy evolution. 

Differentiating Equations~\ref{eq:gam_lt} and \ref{eq:ppar_lt} with respect to LCF time, we find
\begin{align}
	\dv{\gamma'}{t'} =\ & \gamma_u \dv{\gamma}{t'} - u \gamma_u \dv{p_\parallel}{t'} - \gamma_u^2 p_\parallel' \dv{u}{t'} \label{eq:gampdot}\\
	\dv{p_\parallel'}{t'} =\ & \gamma_u \dv{p_\parallel}{t'} - u \gamma_u \dv{\gamma}{t'} - \gamma_u^2 \gamma' \dv{u}{t'} \,.\label{eq:pparpdot}
\end{align}
Setting the change in lab-frame energy to zero and substituting our expressions for the change in the velocity coordinate (Eq.~\ref{eq:udot}) and parallel momentum (Eq.~\ref{eq:ppardot_ad}), we find
\begin{align}
	\eval{\dv{\gamma'}{t'}}_{\rm ad} =\ & -\gamma_u \frac{\gamma'^2 - 1}{\gamma' t} \label{eq:gampdot_ad}\\
	\eval{\dv{p_\parallel'}{t'}}_{\rm ad} =\ & \gamma_u \frac{p_\perp'^2 - u \gamma' p_\parallel'}{\gamma' u t} \,.
\end{align}
In terms of kinetic energy, Equation~\ref{eq:gampdot_ad} is
\begin{equation}
	\eval{\pdv{E'}{t'}}_{\rm ad} = -\gamma_u v'^2 \frac{E' + E_{\rm r}}{t} \,,
\end{equation}
which we can verify by taking appropriate limits. For a non-relativistic fluid, we have $u \ll 1$. In the non-relativistic particle limit $v \ll 1$, and we have $\dd_t E'|_{\rm ad} = -E_{\rm r} v'^2/t = -2 E'/t$, which gives the standard result $E'|_{\rm ad} \propto t^{-2} \propto \rho^{2/3}$. In the ultra-relativistic particle limit $v \simeq 1$, and we have $\dd_t E'|_{\rm ad} = -E'/t$, which gives the standard result $E'|_{\rm ad} \propto t^{-1} \propto \rho^{1/3}$.

The LCF energy also evolves due to collisional losses $\dd_t \gamma'|_{\rm coll}$. Collisional losses preserve direction, so the parallel momentum also evolves to maintain a constant ratio of $p_\perp' / p_\parallel'$. Fixing $p_\perp' / p_\parallel'$ and differentiating Equation~\ref{eq:einstein}, we find
\begin{equation}
	\eval{\dv{p_\parallel'}{t'}}_{\rm coll} = \frac{\gamma' p_\parallel'}{\gamma'^2 - 1} \eval{\dv{\gamma'}{t'}}_{\rm coll} \,.
\end{equation}
Combining adiabatic and collisional terms, the evolution equations are
\begin{align}
	\dv{\gamma'}{t'} =\ & \eval{\dv{\gamma'}{t'}}_{\rm coll} - \gamma_u \frac{\gamma'^2 - 1}{\gamma' t} \label{eq:gampdot}\\
	\dv{p_\parallel'}{t'} =\ & \frac{\gamma' p_\parallel'}{\gamma'^2 - 1} \eval{\dv{\gamma'}{t'}}_{\rm coll} + \gamma_u \frac{p_\perp'^2 - u \gamma' p_\parallel'}{\gamma' u t} \,.\label{eq:pparpdot}
\end{align}

In the trapped limit, we assume the velocity coordinate does not evolve, so $\dd u / \dd t = 0$. In this case, the adiabatic energy loss term is the standard result for an isotropic gas, which is equivalent to Equation~\ref{eq:gampdot_ad}, so the resulting energy evolution equation is equivalent to Equation~\ref{eq:gampdot}.

\subsection{Monte Carlo methods}

We model $\beta$-particles as discrete packets of equal energy in the LCF with state variables of time, velocity coordinate, LCF energy, and LCF parallel momentum. For the radial field scenario, the evolution equations are Equations~\ref{eq:udot}, \ref{eq:gampdot}, and \ref{eq:pparpdot}. For the trapped scenario, $u$ does not evolve and the system is independent of $p_\parallel$.

We assume that the ejecta has a power-law density profile $\rho = C_\rho u^{-\alpha} t^{-3}$, where $C_\rho = (3-\alpha)/(4\pi) M_{\rm ej} v_{\rm ej}^{-(3-\alpha)}$ is a normalization constant. To spawn the packets, we sample the initial time from the energy injection rate $t\sim \partial_t \varepsilon(t)$, the initial velocity coordinate from the ejecta mass profile $u \sim u^2 \rho(u)$, the initial LCF energy from the decay energy spectrum $E' \sim E'\partial_{E'}\Gamma_{\rm dec}(t, E')$, and the initial LCF cosine angle from an isotropic distribution $\mu' \sim \mathcal{U}$. The Lorentz factor and parallel momentum are then $\gamma' = 1 + E'/E_{\rm r}$ and $p_\parallel' = \mu' \sqrt{\gamma^2 - 1}$.

We sample the energy injection rate numerically by using the cumulative trapezoid method to calculate the CDF and invert it with linear interpolation. We cut off the injection rate at early times $<1~{\rm day}$ when most particles thermalize locally and late times $>1~{\rm yr}$ when most particles escape. The late-time cutoff only misses $4\%$ of the decay energy after $1~{\rm day}$. Using the initial time, we select a decay energy spectrum from a pre-tabulated grid of times and sample the initial LCF energy numerically. We sample the velocity coordinate and cosine angle direction using the analytic inverse CDFs of their respective distributions: $u = \xi^{1/(3-\alpha)} v_{\rm ej}$ and $\mu' = 2\xi -1$. 

The mean free path of a packet is given by the energy-dependent transport opacity $\lambda_{\rm mfp} = 1/(\rho \kappa_{\rm t})$. We assume that a packet scatters isotropically in the LCF after reaching an optical depth sampled from an exponential distribution $\tau_* = -\ln \xi$, where the optical depth is defined by $\dd \tau = \kappa_{\rm t} \rho' v' \dd t'$. The LCF energy is conserved during each scattering event, although the lab-frame energy may change. Within each Monte Carlo step, we integrate lab-frame time, velocity coordinate, energy, and parallel momentum as a function of optical depth by multiplying their derivatives with respect to LCF time by $\dd_\tau t'$. We also integrate the energy deposition and number of ionizations for each target, for use in our later post-processing calculation of the ejecta temperature and ionization structure (Sec.~\ref{sec:ionstate}).

We set the integration step $\Delta \tau$ to the minimum of $\tau_* - \tau$, $f$, $f \cdot E'/\abs{\dd_\tau E'}$, and $f \cdot R_{\rm ej} / (v' \dd_\tau t')$, where we use $f = 0.01$. This choice ensures that we never overshoot $\tau_*$ and that changes in optical depth, energy, and velocity coordinate are well sampled. At each integration step, we compute the velocity coordinate and LCF energy, and we select the stopping power, transport opacity, and ionization rates from a pre-tabulated grid of energies. The stopping power and transport opacity include contributions from all interactions and targets, as described in Section~\ref{sec:total_xsec} (see Eqs.~\ref{eq:Stot} and \ref{eq:sigttot}).

If $\tau$ reaches $\tau_*$, we model isotropic scattering by resampling the LCF cosine angle from a uniform distribution. However, if an end condition is reached, we terminate the packet evolution. The end conditions are (i) $E < E_{\rm thm} = 1~{\rm eV}$, complete thermalization; (ii) $u > v_{\rm ej}$, escape; and (iii) $t > t_{\rm max} = 1~{\rm yr}$, persistence. We assume that persistent packets eventually escape with negligible additional thermalization because the ejecta becomes very diffuse at late times.

Each packet represents an LCF energy $E'_{\rm tot}/N_{\rm pac}$ or, equivalently, a number of particles $E'_{\rm tot}/(E_{\rm i} N_{\rm pac})$, where $E'_{\rm tot} = M_{\rm ej} \int \dd t Q_{\rm inj}$ is the total injected LCF energy, $N_{\rm pac}$ is the total number of packets, and $E_{\rm i}'$ is the LCF energy of the packet at injection. As the simulation runs, we keep track of energy deposition and ionizations on a $128^2$ grid in logarithmic time and velocity coordinate. 

We neglect the effect of changes in the elemental abundances over time, which has a negligible effect on the stopping power and transport opacity, instead fixing the elemental abundances to their value at 30 days post-merger. Similarly, we neglect the effect of changes in the Coulomb logarithm over time (Eq.~\ref{eq:coulog}), instead setting it using our nominal temperature of $10^3~{\rm K}$ and the average ejecta density at 30 days post-merger.

We implement these Monte Carlo simulations in a \textsc{Python} script, parallelized with \textsc{mpi4py} \citep{Dalcin+2005} and accelerated by \textsc{numba} \citep{Lam+2015} using just-in-time compilation. We run simulations over a variety of ejecta models, parameterized by ejecta mass, velocity, $r$-process model, ionization state, and density profile power-law exponent. For our fiducial ejecta model, we use a strong $r$-process with singly ionized ejecta, $M_{\rm ej} = 0.003~M_\odot$, $v_{\rm ej} = 0.2 c$, and $\alpha=1$. For each transport limit, we run a grid of ejecta mass, velocity, and density power-law slope $\{0.001, 0.003, 0.01, 0.03\}~M_\odot \times \{0.1, 0.2, 0.3\}c \times \{1.0, 2.0\}$ and a grid of $r$-process models and ionization states \{weak, medium, strong\}$\times$\{II, III, IV\}, making 56 simulations in total. Each model was run on eight 2.9 GHz Intel Cascade Lake processors for 12 hours, resulting in roughly 50 million packets.

\section{Monte Carlo Results}
\label{sec:results}
\subsection{Particle trajectories}
\label{sec:traj}

\begin{figure*}
    \centering
    \includegraphics[width=0.8\linewidth]{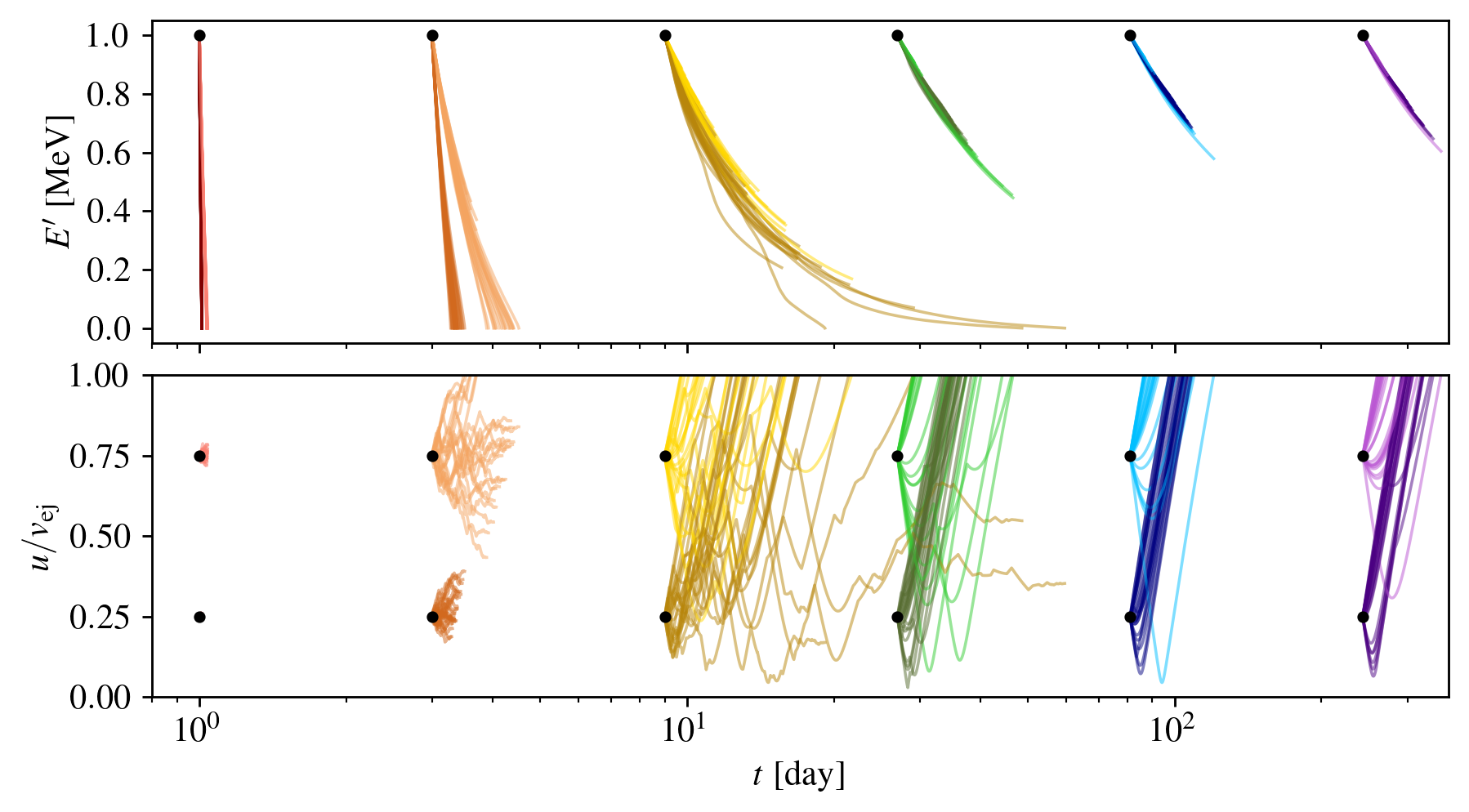}
    \caption{Velocity coordinate $u/v_{\rm ej} = u / (0.2 c)$ (bottom panel) and comoving energy (top panel) as a function of time in our fiducial ejecta model with a radial magnetic field for 1-MeV particles injected by-hand at times 1 (red lines), 3 (orange lines), 9 (yellow lines), 27 (green lines), 81 (blue lines), and 243 (purple lines) days and velocity coordinates $0.05c$ (dark lines) and $0.15c$ (light lines). We inject 24 particles at each injection site (black dots). Initially, particles thermalize locally in space and time (red lines). At intermediate times, particles persist in the ejecta and thermalize non-locally in space and time (orange and yellow lines). Some particles escape before they completely thermalize, while others are trapped by scattering, especially those injected at small velocity coordinates where the density is higher. At late times, particles escape before most of their energy thermalizes (green, blue, and purple lines).}
    \label{fig:traj}
\end{figure*}

The expansion of the ejecta has multiple effects on $\beta$-particle thermalization. As the stopping power decreases, particles persist longer in the ejecta, so thermalization becomes increasingly non-local in space and time, and particles lose more energy to adiabatic expansion. Simultaneously, as the transport cross section decreases, scattering becomes less frequent and particles are more likely to escape before completely thermalizing. In Figure~\ref{fig:traj}, we showcase these effects in our fiducial ejecta model for a radial magnetic field by injecting 1-MeV particles by-hand at times 1, 3, 9, 27, 81, and 243 days and velocity coordinates $0.05c$ and $0.15c$. 

The particles are injected with an isotropic pitch angle distribution in the LCF, which corresponds to a slightly forward-peaked distribution in the lab frame due to relativistic beaming. Therefore, almost half of the particles initially move towards smaller velocity coordinate. At early times $\sim {\rm day}$, particles thermalize locally in space and time. Later on, particles persist in the ejecta and thermalization becomes non-local. 

As the typical distance particles are transported before thermalizing increases, transport becomes increasingly biased towards larger velocity coordinates due to magnetic mirroring and focusing, or due to geometry in the equivalent unmagnetized scenario. In Figure~\ref{fig:traj}, this is most clear in the trajectories of particles injected at 3 days and $0.25 v_{\rm ej}$. However, it also applies at larger velocity coordinate and later times. For particles injected at 3 days and $0.75 v_{\rm ej}$, the furthest transported particles escape and their trajectories are cut off.

Some particles escape before they completely thermalize, while others are trapped by scattering, especially those injected at small velocity coordinate where the density is higher. Initially, particle motion can be treated as a diffusive process with coefficient $D \simeq (1/3) v \lambda_{\rm mfp}$, where $v$ is the particle velocity and $\lambda_{\rm mfp} = 1/(\kappa_{\rm t} \rho) \propto M_{\rm ej}^{-1} v_{\rm ej}^{3} t^{3}$ is the particle mean free path. In this limit, the typical escape time is $t_{\rm esc} \sim R_{\rm ej}^2 / D \propto M_{\rm ej} v_{\rm ej}^{-1} t^{-1}$. 

At late times, scattering becomes inefficient and particles escape before most of their energy thermalizes. At large velocity coordinate, particles that initially move outward escape most efficiently while particles that initially move inward make a significant movement through the inner ejecta before ultimately escaping, allowing them to deposit a greater fraction of their energy.

Particle motion can be treated as free-streaming when $\lambda_{\rm mfp} \gtrsim R_{\rm ej}$, which occurs at time $t_{\rm free} \propto M_{\rm ej}^{1/2} v_{\rm ej}^{-1}$. In this limit, the typical escape time is $t_{\rm esc} \sim R_{\rm ej} / v \propto v_{\rm ej} t$, which has the same time dependence as the timescale for adiabatic losses $t_{\rm ad} = E / \dot{E}|_{\rm ad} \propto t$. These estimates demonstrate that particle escape becomes increasingly important relative to adiabatic losses during the diffusive transport phase, but stalls relative to adiabatic losses during the free-streaming transport phase.

\subsection{Thermalization efficiency and the role of transport}

\begin{figure*}
    \centering
    \includegraphics[width=1.0\linewidth]{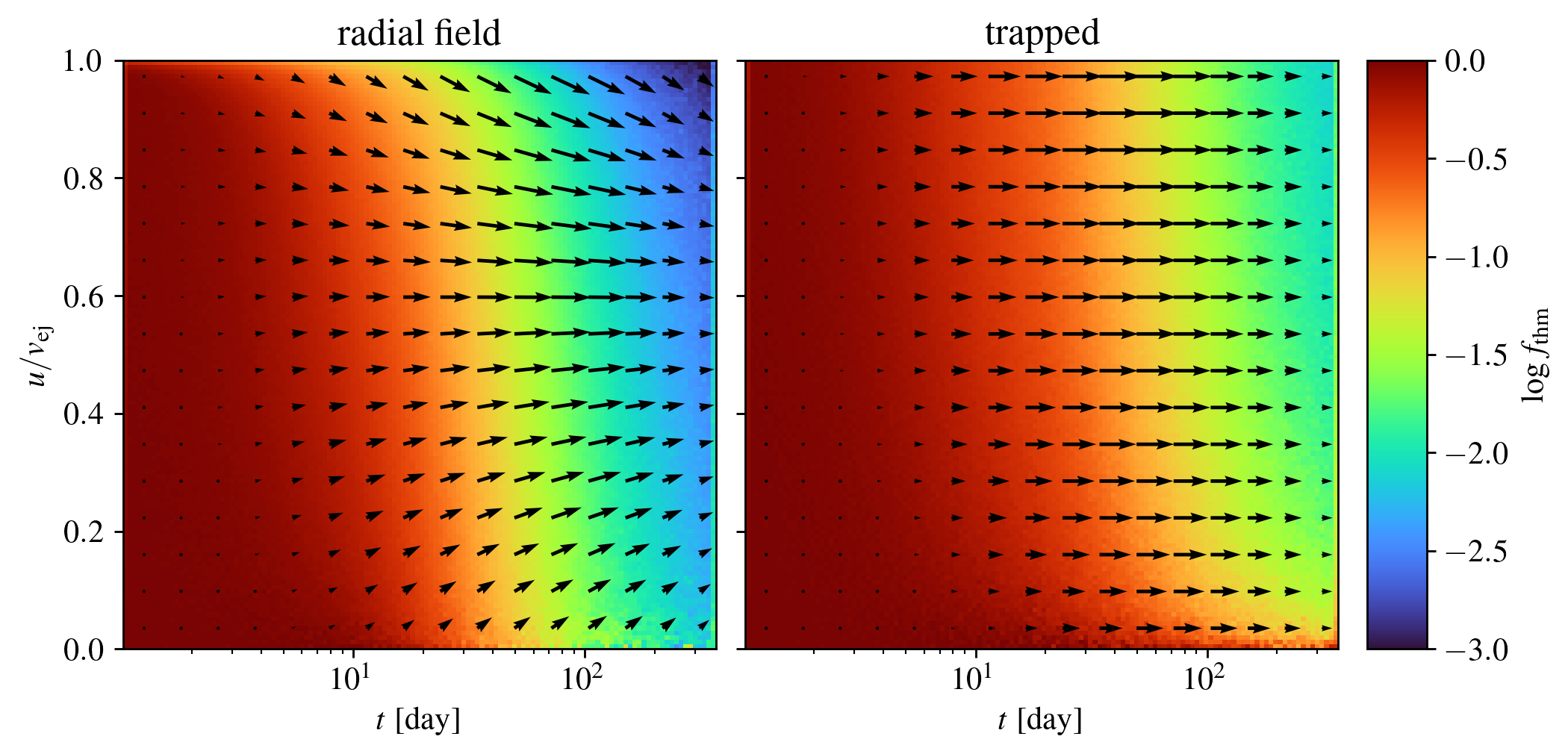}
    \caption{The thermalization efficiency at each bin in velocity coordinate and time in our fiducial ejecta model in the radial magnetic field (left panel) and trapped (right panel) limits. The arrows show the average time and velocity-coordinate displacement between where energy is injected (arrow base) and where it thermalizes. Arrow lengths are re-normalized for visual clarity. Initially, the dense ejecta promotes local thermalization in space and time for both transport scenarios. In the trapped limit, the thermalization efficiency decreases independently at each velocity coordinate, with a more rapid decrease in the diffuse outer ejecta. In the radial field limit, particles injected at small velocity coordinates move outwards before depositing their energy. At large velocity coordinate, outward-moving particles escape, so only inward-moving particles contribute to thermalization and deposit their energy. These effects lower the thermalization efficiency and make it more uniform across velocity coordinates in the radial field scenario.}
    \label{fig:mc2d}
\end{figure*}

At each bin in time and velocity coordinate, we define the thermalization efficiency $f_{\rm thm}$ as the ratio of thermalized to injected energy in that bin, where the thermalized energy is defined as any energy deposited from packet to plasma by any microphysical process. Energy carried by secondaries is assumed to thermalize locally. At a given velocity coordinate and time, the thermalized energy includes contributions from particles injected at earlier times and different velocity coordinates. In Figure~\ref{fig:mc2d}, we show the thermalization efficiency for our fiducial ejecta model in the radial field and trapped limits. We use arrows to show the average time and velocity coordinate displacement between where energy is injected and where it thermalizes.

Initially, the dense ejecta promotes local thermalization in space and time for both transport scenarios. In the trapped limit, the thermalization efficiency decreases independently at each velocity coordinate, with a more rapid decrease in the diffuse outer ejecta. In the radial field limit, particles injected at small velocity coordinates move outwards before depositing their energy. At large velocity coordinate, outward-moving particles escape, so only inward-moving particles contribute to thermalization and deposit their energy. These effects lower the thermalization efficiency and make it more uniform across velocity coordinates in the radial field scenario.

\begin{figure*}
    \centering
    \includegraphics[width=0.67\linewidth]{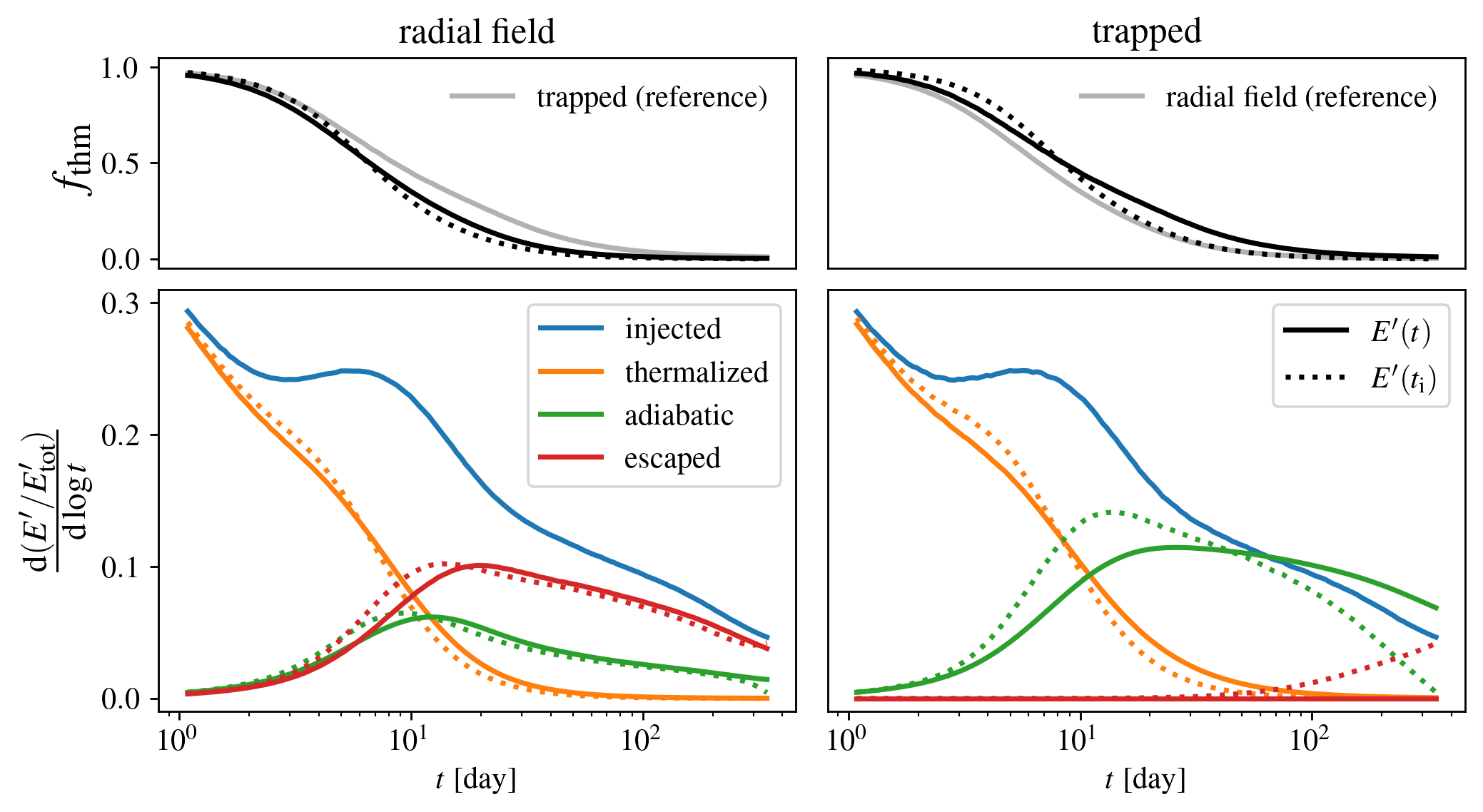}
    \caption{Thermalization efficiency (top panels), injected decay energy (bottom panels, blue lines), and energy losses (bottom panels) due to thermalization (orange lines), adiabatic losses (green lines), and escape (red lines) as a function of time in our fiducial ejecta model in the radial magnetic field (left panels) and trapped (right panels) limits. In the top panels, we show the thermalization efficiency for the other transport limit as a faded black line for reference. Solid lines show quantities as a function of the time at which energy is deposited or lost, while dotted lines show quantities as a function of the time at which the particles carrying that energy were originally injected. At a given time, the injected energy is equal to the sum of the injection-time energy losses. The thermalization efficiency decreases over time as the ejecta becomes more diffuse. In the trapped limit, this decrease is driven by adiabatic losses. In the radial field limit, this decrease is mostly driven by particle escape, although there are still significant adiabatic losses.}
    \label{fig:ethmt}
\end{figure*}

In Figure~\ref{fig:ethmt}, we show the thermalization efficiency integrated over velocity coordinate. We also show the injected decay energy and the energy losses due to thermalization, adiabatic losses, and escape. We plot quantities as a function of the time at which the energy is deposited or lost, and as a function of the time at which the particles carrying that energy were originally injected. At a given time, the injected energy is equal to the sum of the injection-time energy losses. The deposition-time and injection-time thermalization efficiencies are equal to the ratios of the corresponding thermalized energy to the injected energy.

As an example of how to interpret this plot, consider $t=10~{\rm days}$. The blue line represents the $\beta$-decay energy injected at $10~{\rm days}$. The dotted lines represent the portions of that injected energy that ultimately thermalize, escape, or get lost to adiabatic expansion. In the radial field limit, energy is split between these channels in roughly equal portions, while in the trapped limit, energy is only split between thermalization and adiabatic losses. The solid lines represent the energy that actually thermalizes, escapes, or gets lost to adiabatic expansion at $10~{\rm days}$, including contributions from particles injected at earlier times. For example, in the radial-field limit, the thermalized energy at $10~{\rm days}$ (solid orange) is larger than the eventual thermalization of energy injected at $10~{\rm days}$ (dotted orange).

The thermalization efficiency decreases over time as the ejecta becomes more diffuse. In the trapped limit, this behavior is driven by adiabatic losses as particles persist longer in the ejecta. In the radial field limit, this behavior is mostly driven by particle escape as transport becomes more efficient, although there are still significant adiabatic losses. Energy losses due to escape initially increase faster than those due to adiabatic expansion, but follow adiabatic losses at late times. The transition occurs around 20 days, consistent with the transition from random walk to free-streaming particle trajectories in Figure~\ref{fig:traj}.

During the initial phase of local thermalization, energy deposition closely tracks local injection and the deposition-time thermalization losses lie below the injection-time losses. At late times, particles persist before thermalizing, and the energy deposited at a given time becomes dominated by particles injected earlier. As a result, the deposition-time thermalization losses exceed the injection-time losses. In the trapped limit, the energy escaping at each time is identically zero because particles cannot physically escape the ejecta. However, the injection-time escape losses rise at late times as an increasing fraction of injected particles persist to the end of the simulation.

\begin{figure*}
    \centering
    \includegraphics[width=0.67\linewidth]{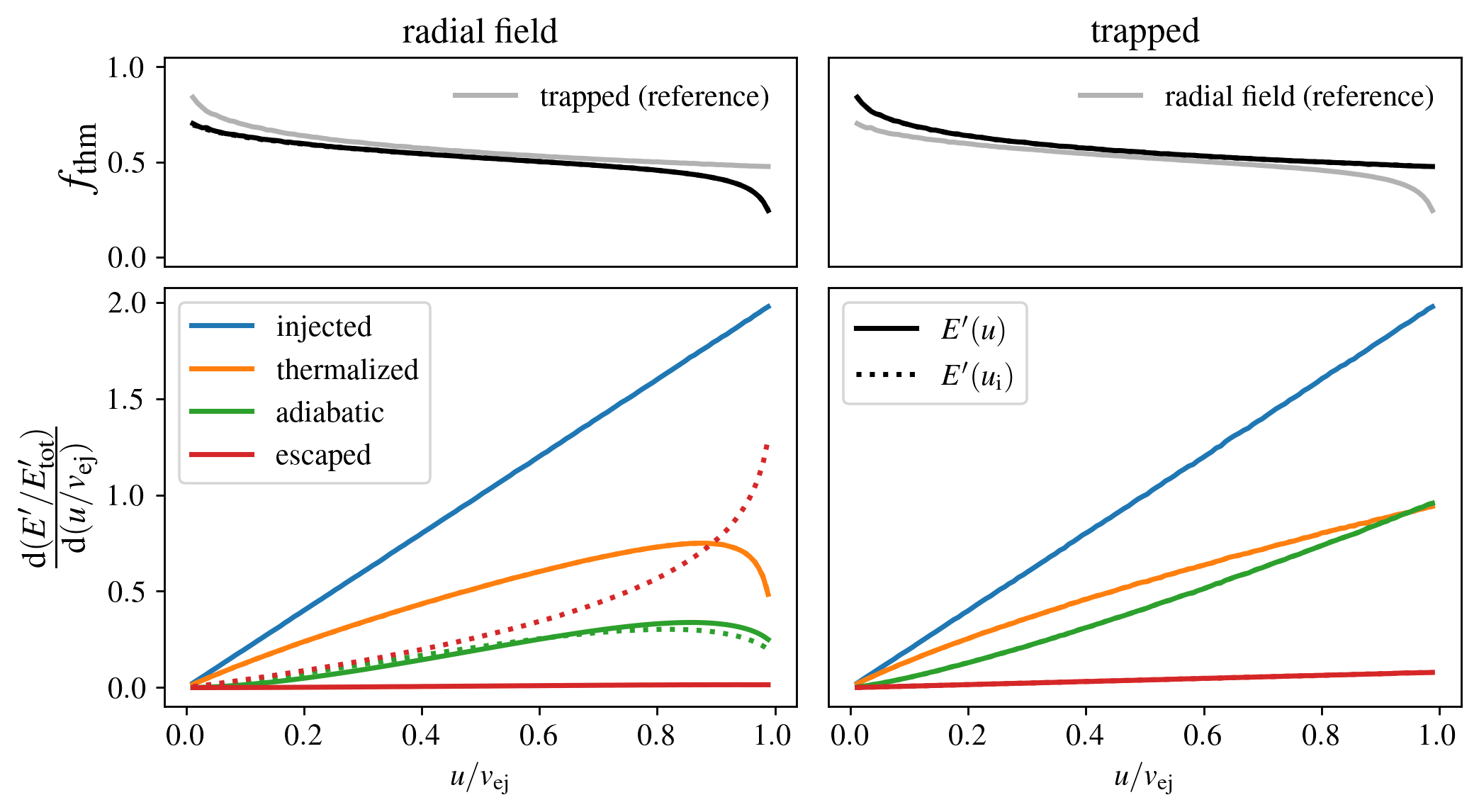}
    \caption{Same as Figure~\ref{fig:ethmt}, but as a function of velocity coordinate instead of time. The energy escaping at each velocity coordinate is due to particles that persist until the end of the simulation (solid red lines). The injected energy increases linearly with velocity coordinate because $\dd M / \dd u \propto u$ for our fiducial ejecta model with $\alpha = 1$. The thermalization efficiency decreases with increasing velocity coordinate as the ejecta becomes more diffuse. In the trapped limit, this decrease is driven by adiabatic losses. In the radial field limit, this decrease is driven by particle escape, although there are still significant adiabatic losses. The radial field scenario has lower thermalization efficiency at every velocity coordinate, with the largest differences in the innermost and outermost ejecta. }
    \label{fig:ethmu}
\end{figure*}

In Figure~\ref{fig:ethmu}, we integrate the thermalization efficiency over time instead of velocity coordinate. We also show the injected decay energy and energy losses, similar to Figure~\ref{fig:ethmt}. The energy escaping at each velocity coordinate is due to particles that persist until the end of the simulation. In the radial field limit, $\lesssim 0.9\%$ of the decay energy persists to the end of the simulation across all velocity coordinates, while in the trapped limit, $3-4\%$ of decay energy persists across all velocity coordinates.

The results are similar to Figure~\ref{fig:ethmt}. The thermalization efficiency decreases with increasing velocity coordinate as the ejecta becomes more diffuse. In the trapped limit, this decrease is driven by adiabatic losses. In the radial field limit, this decrease is driven by particle escape, although there are still significant adiabatic losses. The radial field limit has lower thermalization efficiency at every velocity coordinate, with the largest differences in the innermost and outermost ejecta. At each velocity coordinate, most of the energy deposition is contributed by early times when energy deposition closely tracks local injection, so deposition-coordinate and injection-coordinate thermalization losses are nearly identical.

\subsection{Inefficiency time and asymptotic slope}
\label{sec:ineff_time}

Motivated by the analytic arguments in previous works \citep{Barnes+2016, Waxman+2019, Kasen&Barnes2019, Shenhar+2024}, we fit our time-dependent thermalization efficiencies with an analytic form,
\begin{equation}
	f_{\rm thm}(t) = \frac{\ln(1 + \tau)}{\tau}, \quad \tau \equiv 2 \left(\frac{t}{t_{{\rm ineff}}}\right)^{n} \,,
	\label{eq:anal_fit}
\end{equation}
where the inefficiency time $t_{\rm ineff}$ and the asymptotic slope $n$ are free parameters. The inefficiency time represents the time after which thermalization starts to become inefficient. This formula is an \textit{ad hoc} interpolation between limiting behaviors, with $f_{\rm thm} \simeq 1$ for $t \ll t_{\rm ineff}$ and approximately $f_{\rm thm} \propto t^{-n}$ for $t \gg t_{\rm ineff}$, with a logarithmic correction. This form produces a better fit to our simulations than the alternative form $f_{\rm thm}(t) = [1 + (t/t_{\rm ineff})]^{-n}$ derived by \citet{Kasen&Barnes2019}.

Thermalization becomes inefficient when the timescale for thermalization $t_{\rm thm} \approx E / \dot{E}|_{\rm coll} \propto \rho^{-1} \propto M_{\rm ej}^{-1} v_{\rm ej}^3 (u/v_{\rm ej})^\alpha t^{3}$ becomes longer than the timescale for adiabatic expansion $t_{\rm exp} = t$ \citep{Barnes+2016}. Setting $t_{\rm exp} = t_{\rm thm}$ and solving for $t$, we find
\begin{equation}
	t_{\rm ineff} \propto M_{\rm ej}^{1/2} v_{\rm ej}^{-3/2} (u/v_{\rm ej})^{-\alpha/2} \,.
	\label{eq:teff}
\end{equation}

For the asymptotic slope, \citet{Barnes+2016} derived a value $n=2$ for an idealized model where collisional energy losses are independent of energy, the decay spectrum is independent of time, and the energy injection rate declines as $Q_{\rm inj}\propto t^{-1}$. For a more general model, the asymptotic slope can deviate from this value \citep{Kasen&Barnes2019}.

\begin{figure*}
    \centering
    \includegraphics[width=1.0\linewidth]{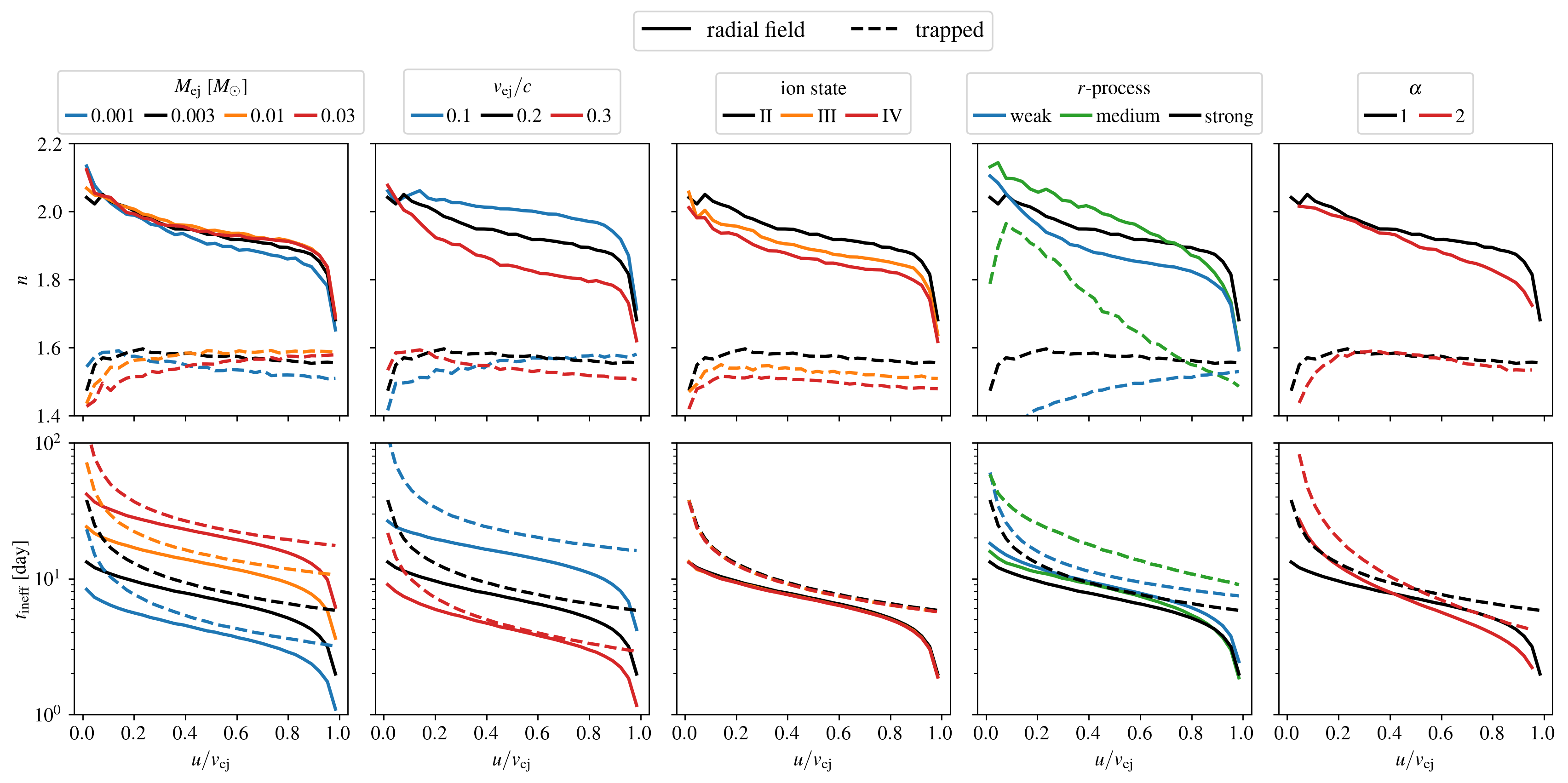}
    \caption{The inefficiency time (bottom panels) and asymptotic slope (top panels) as a function of velocity coordinate, derived by fitting Equation~\ref{eq:anal_fit} to the thermalization efficiency at each velocity coordinate. In each column, we vary a different ejecta model parameter---ejecta mass (first column), velocity (second column), ionization state (third column), $r$-process abundance pattern (fourth column), and density profile power-law exponent (fifth column). Each curve is colored by its parameter value in the color bar above the column, with the fiducial ejecta model in black. We show curves in radial magnetic field (solid lines) and trapped (dashed lines) limits. In the trapped limit, the inefficiency time scales with velocity coordinate as $t_{\rm ineff} \propto u^{-1/2}$, as expected from Equation~\ref{eq:teff} with $\alpha=1$. In the radial field limit, the inefficiency time is shorter at all velocity coordinates, especially in the innermost and outermost ejecta due to particle transport and escape. }
    \label{fig:ej_mod}
\end{figure*}

We divide the ejecta into 32 equal bins in velocity coordinate and fit Equation~\ref{eq:anal_fit} to the thermalization efficiency integrated across each bin. The integration is necessary to reduce numerical noise. We perform the curve fit in log-log space using least-squares regression as implemented in \texttt{scipy.optimize.curve\_fit}. In Figure~\ref{fig:ej_mod}, we show the inefficiency time and asymptotic slope as a function of velocity coordinate for many ejecta models, varying each ejecta model parameter separately. We also fit Equation~\ref{eq:anal_fit} to the thermalization efficiency integrated across all velocity coordinates to get a simplified description of the global thermalization. In this work, we are mostly interested in the spatially dependent thermalization efficiency, but we tabulate the results in Appendix~\ref{app:fit} for completeness. 

We show the fiducial ejecta model in black. The typical inefficiency time of $6~{\rm days}$ is broadly consistent with thermalization timescales reported in previous work \citep{Metzger+2010, Barnes+2016, Waxman+2019, Kasen&Barnes2019, Hotokezaka&Nakar2020, Shenhar+2024}. In the trapped limit, the inefficiency time scales with velocity coordinate as $t_{\rm ineff} \propto u^{-1/2}$, as expected from Equation~\ref{eq:teff} with $\alpha=1$. In the radial field limit, the inefficiency time is shorter at all velocity coordinates, especially in the innermost and outermost ejecta. In the innermost ejecta, this behavior occurs because transport is biased towards larger velocity coordinate where thermalization is less efficient. In the outermost ejecta, this behavior occurs because particles efficiently escape the ejecta. 

In the trapped limit, the asymptotic slope is nearly constant across velocity coordinates with a value $n \simeq 1.6$. In the radial field limit, the asymptotic slope is larger, with values $n \simeq 1.9$--2.0, due to increasing particle escape at late times. Both limits are consistent with the model of \citet{Waxman+2019}. In the trapped limit, their model predicts that the deposited energy scales as roughly $\propto t^{-2.8}$, independent of $Q_{\rm inj}$. For our $Q_{\rm inj} \propto t^{-1.3}$ (Eq.~\ref{eq:inj_rate}), this corresponds to $n = 1.5$. In the limit of free escape, which is equivalent to the radial field limit in our 1D set-up (Sec.~\ref{sec:transport_equations}),  their model predicts $n = 2$. We also find a smaller asymptotic slope in the outer ejecta where particle escape is most efficient.

\subsection{Dependence on the ejecta model}
\label{sec:ej_mod}

The inefficiency time increases with ejecta mass and decreases with ejecta velocity for both transport limits, as expected from Equation~\ref{eq:teff}. The difference in inefficiency time between the transport limits shrinks for larger ejecta velocities because of changes to the escape timescale. At larger ejecta velocity, the transition to free-streaming occurs earlier ($t_{\rm free}\propto M_{\rm ej}^{1/2} v_{\rm ej}^{-1}$) and the escape time in that phase increases ($t_{\rm esc} \propto v_{\rm ej} t$) because particles must cover a large distance to escape (Sec.~\ref{sec:traj}). As a consequence, particles injected at late times persist longer in more rapidly expanding ejecta, enhancing the late-time thermalization efficiency for those ejecta models, although this effect is partially counter-balanced by greater relativistic focusing. This effect also makes the asymptotic slope more shallow at larger ejecta velocity.

The inefficiency time is insensitive to the ionization state of the ejecta. However, the asymptotic slope tends to decrease at higher ionization states due to the higher free electron densities in more ionized ejecta, which enhances the thermalization losses of lower energy $\beta$-particles through M{\o}ller scattering (Fig.~\ref{fig:stop_kap}). In the context of the \citet{Waxman+2019} model, this can be understood as a change in the effective power-law slope of the stopping power.

The $r$-process abundance pattern has three effects. First, the abundance pattern changes the stopping power and transport cross section. These quantities are similar for the strong and medium $r$-process abundances, but are reduced by a factor of a few for the weak $r$-process due to the smaller ionization losses associated with elements from the first $r$-process peak.

Second, the abundance pattern changes the time-dependent injection rate (Fig.~\ref{fig:rate}). The injection rate oscillates about an overall power law trend depending on the detailed decay pattern. The timing of these oscillations near the transition to inefficient thermalization has a significant effect on the thermalization efficiency. Relative to the strong $r$-process injection rate, the medium $r$-process rate declines more rapidly from 6--30 days and more slowly at $\gtrsim 30$ days, while the weak $r$-process rate declines more rapidly from 3--20 days and more slowly from 20--100 days.

Third, the abundance pattern changes the decay spectrum (Fig.~\ref{fig:decspec}). At typical inefficiency times in our simulations of $t_{\rm ineff} \simeq 1$--10 days, the average decay electron produced by the strong, medium, and weak $r$-process has an energy around 250~keV, 400~keV, and 200~keV, respectively (Fig.~\ref{fig:rate}). 

In the trapped limit, the medium $r$-process has the longest inefficiency time. This behavior is driven by the more rapid decline in the injection rate at intermediate times, which enhances the thermalization efficiency because delayed deposition from particles injected earlier, when the injection rate was higher, contributes disproportionately relative to the lower instantaneous injection rate. This effect is partially counterbalanced by the more energetic decay spectrum relative to the strong $r$-process, which makes thermalization less efficient because the stopping power is lower at higher energies and particles persist in the ejecta for longer durations.

Similarly, the weak $r$-process also has a longer inefficiency time. However, in the weak $r$-process case, the effect is additionally counterbalanced by the reduction in stopping power and transport opacity from the atomic physics. The net result is an inefficiency time in between those for the medium and strong $r$-processes.

In the radial field limit, the inefficiency times for the medium and weak $r$-processes are closer to the strong $r$-process value, suggesting that the effects of the abundance pattern are diluted by transport. In particular, the effect of fluctuations in the injection rate may be weaker because particles move to larger velocity coordinate as they thermalize, where for $\alpha=1$ they make up a smaller fraction of the injected energy. 

For a steeper density power-law index $\alpha = 2$ in the trapped limit, the inefficiency time declines more steeply with velocity coordinate as $t_{\rm ineff} \propto u^{-1}$, following Equation~\ref{eq:teff}. For both transport scenarios, thermalization is more efficient because a greater fraction of the energy is injected at smaller radii where the inefficiency time is the longest. 

Even with a relatively limited set of ejecta models, the diversity of thermalization behaviors suggests that results from a single ejecta model should not be interpreted as universal. Aside from the ejecta mass and velocity, thermalization is also sensitive to fluctuations in the energy injection rate relative to the overall power law trend and the density profile of the ejecta. In comparison, thermalization is less sensitive to the ejecta ionization state, so energy deposition can be reasonably modeled without computing the ionization state self-consistently.

\subsection{Analytic prescription}
\label{sec:analytic_prescription}

Using the same curve-fitting procedure described in Section~\ref{sec:ineff_time}, we fit the spatially dependent thermalization efficiency across our series in ejecta mass and velocity with the following equations:
\begin{align}
	f_{\rm thm} =\ & \ln(1+\tau) / \tau, \quad \tau \equiv 2 (t/t_{\rm ineff})^{n} \label{eq:presfirst}\\
	t_{\rm ineff}^{({\rm trp})} =\ & t_{\rm ineff, 0} M_{0.003}^{1/2} v_{0.2}^{-3/2} x^{-\alpha/2}\\
	t_{\rm ineff}^{({\rm rad})} =\ & v_{0.2}^{1/3} b(x; x_{\rm p}, r_{\rm p}, 2/3) t_{\rm ineff}^{({\rm trp})} \,,\label{eq:preslast}
\end{align}
where $M_{0.003} \equiv M_{\rm ej}/0.003~M_\odot$, $v_{0.2} \equiv v_{\rm ej} / 0.2c$, $x \equiv u/v_{\rm ej}$, the superscripts denote the radial field (rad) and trapped (trp) limits, and $b(x)$ is a normalized beta-profile given by
\begin{equation}
	b(x; x_{\rm p}, r_{\rm p}, w) \equiv r_{\rm p} \left( \frac{x}{x_{\rm p}}\right)^{w x_{\rm p}} \left( \frac{1 - x}{1 - x_{\rm p}}\right)^{w (1 - x_{\rm p})} \,.
\end{equation}
The profile peaks at $(x_{\rm p}, r_{\rm p})$ and its width is controlled by $w$, which we fix to $w=2/3$ since it provides a reasonable fit across all models. 

This choice captures the expected behavior in the radial field limit. Near $x=0$, the inefficiency time should trend towards zero as the volume of the corresponding ejecta shell decreases and magnetic focusing becomes stronger. Near $x=1$, the inefficiency time should also trend towards zero as particles become increasingly likely to escape. These edge effects are weaker near the peak of the profile $x \simeq x_{\rm p}$, so the inefficiency time should be closest to its value in the trapped limit. The asymptotic slope is roughly independent of ejecta mass and velocity in the trapped limit, and can be approximated by a linear dependence on ejecta velocity in the radial field limit.

\begin{table}
    \centering
    \caption{Fit parameters for Equations~\ref{eq:presfirst}--\ref{eq:preslast}, which describe the thermalization efficiency as a function of time, velocity coordinate (position), ejecta mass, and ejecta velocity. Each row shows the results for models with a different density power-law slope $\alpha$. }
    \label{tab:anal_fit}
    \begin{tabular}{cccccc}
        \hline
        $\alpha$ &
        $t_{\rm ineff,0}^{(\rm trp)}$ [day] &
        $x_{\rm p}$ &
        $r_{\rm p}$ &
        $n^{(\rm trp)}$ &
        $n^{(\rm rad)}$ \\
        \hline
        $1$ & $5.37$ & $0.53$ & $0.91$ & $1.54$ &
        $2.08 - 0.15 (v_{\rm ej}/0.2c)$ \\
        $2$ & $3.96$ & $0.59$ & $0.80$ & $1.53$ &
        $2.07 - 0.16 (v_{\rm ej}/0.2c)$ \\
        \hline
    \end{tabular}
\end{table}

\begin{figure}
    \centering
    \includegraphics[width=1.0\linewidth]{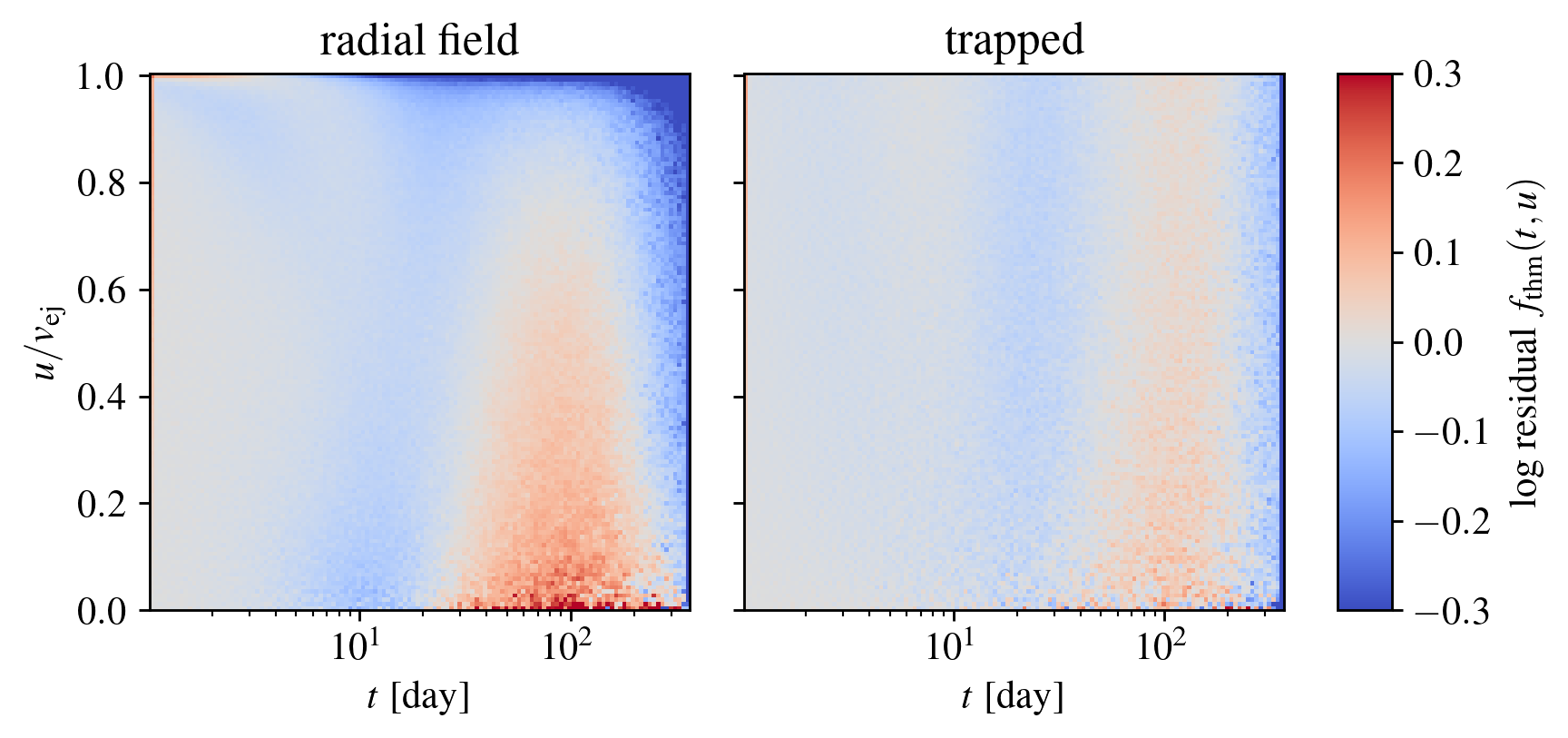}
    \caption{Thermalization efficiency logarithmic residuals ${\rm log} (f_{\rm thm}^{({\rm analytic})} / f_{\rm thm}^{({\rm mc})})$ for the analytic prescription described by Equations~\ref{eq:presfirst}-\ref{eq:preslast} at each bin in velocity coordinate and time for the fiducial ejecta model in the radial-field (left panel) and trapped (right panel) limits. The residuals for all ejecta models are given in Figure~\ref{fig:res_all}. In the trapped limit, all models have logarithmic residuals $\lesssim 0.1~{\rm dex}$ at all locations and times. In the radial field limit, the largest residuals $\sim 0.3~{\rm dex}$ occur in the innermost and outermost ejecta at late times, mostly because we approximate the asymptotic slope as independent of velocity coordinate. These results indicate that our prescription is a reasonable approximation across a range of ejecta conditions.}
    \label{fig:res}
\end{figure}

In Table~\ref{tab:anal_fit}, we show the resulting fit parameters for models with density power-law slopes of $\alpha=1$ and $\alpha=2$. To evaluate the accuracy of this prescription, we compute the logarithmic residuals relative to our Monte Carlo results ${\rm log} (f_{\rm thm}^{({\rm analytic})} / f_{\rm thm}^{({\rm mc})})$. Across all locations, times, and ejecta models, the median absolute residuals are $0.036$ and $0.019$ dex in the radial field and trapped limits respectively. In Figure~\ref{fig:res}, we plot the residuals at each location and time for the fiducial ejecta model. The residuals for all ejecta models are shown in Figure~\ref{fig:res_all}. In the trapped limit, all models have residuals $\lesssim 0.1~{\rm dex}$ at all locations and times. In the radial field limit, the largest residuals $\sim 0.3~{\rm dex}$ occur in the innermost and outermost ejecta at late times, mostly because we approximate the asymptotic slope as independent of velocity coordinate.

This prescription provides a simple, computationally cheap approximation for the spatially dependent thermalization efficiency that can be incorporated into semi-analytic and/or radiative-transfer KN emission models. However, it should only be used as a first approximation since, in detail, the thermalization efficiency depends on the time-dependent nuclear decay rate. In general, we recommend explicit charged particle transport whenever feasible, particularly in multi-dimensional simulations where magnetic field geometry and ejecta structure may differ substantially from the idealized models considered here. We provide $\beta$-particle stopping and transport cross sections for use in explicit transport models (see the \hyperref[sec:data]{Data Availability statement}).

\section{Ejecta ionization structure}
\label{sec:ionstate}
The ejecta temperature evolves according to the first law of thermodynamics \citep{Jerkstrand2011, Pognan+2022}, i.e.
\begin{equation}
	\dv{T}{t} = \frac{\mathcal{H} - \mathcal{C}}{\frac{3}{2} k_{\rm B} n} - \frac{2 T}{t} - \frac{T}{1 + x_{\rm e, thm}} \dv{x_{\rm e, thm}}{t} \,,
	\label{eq:tempdot}
\end{equation}
where $\mathcal{H} = f_{\rm thm} \rho Q_{\rm inj}$ is the volumetric heating rate, $\mathcal{C}$ is the volumetric cooling rate, $n = n_{\rm i} + n_{\rm e, thm}$ is the total thermal particle number density, and $x_{\rm e, thm} \equiv n_{\rm e, thm} / n_{\rm i}$ is the free thermal electron number fraction. The second term accounts for adiabatic cooling and the third term accounts for changes in the number of thermal electrons. 

In practice, the electron number fraction changes slowly relative to the expansion timescale, so the ionization term is always subdominant to the adiabatic term and we neglect it. Cooling in the nebular phase is dominated by line cooling of heavy $r$-process elements \citep{Pognan+2022}. Since this quantity is proportional to $n_{\rm e} n_{\rm i}$, we define $\mathcal{C} \equiv n_{\rm e} n_{\rm i} \Lambda$, where $\Lambda$ is independent of density for sufficiently low densities, i.e. $n \lesssim 10^4~{\rm cm^{-3}}$ \citep{Hotokezaka+2021}. 

We do not model the radiation field, and instead assume that all line photons escape locally. This approximation is reasonable in the nebular phase when the ejecta are optically thin. However, we may underestimate the temperature at early times when the radiation is partially coupled to the gas.

The ionization fractions evolve according to the rate equations \citep{Pognan+2022}
\begin{equation}
	\dv{x_{i, q}}{t} = \Gamma_{i, q-1} x_{i, q-1} - (\alpha_{i, q} n_{\rm e, thm} + \Gamma_{i, q}) x_{i, q} + \alpha_{i, q+1} n_{\rm e, thm} x_{i, q+1} \,,
	\label{eq:xijdot}
\end{equation}
where $\Gamma_{i, q}$ and $\alpha_{i, q}$ are the ionization rate and recombination rate coefficient for species $(i, q)$ and $n_{\rm e, thm}$ is the number density of free thermal electrons given by
\begin{equation}
	n_{\rm e, thm} = \rho N_{\rm A} \sum_i Y_i \sum_q x_{i q} q \,.
\end{equation}
The total cooling function is the sum of contributions from each species,
\begin{equation}
	\Lambda = \mu_{\rm mmw} \sum_i Y_i \sum_q x_{i q} \Lambda_{i q} \,.
\end{equation}

For $r$-process elements, dielectronic recombination (DR) often dominates over radiative recombination (RR), sometimes by more than an order-of-magnitude \citep{Hotokezaka+2021, Singh+2025}. At typical nebular temperatures $10^{3-4}~{\rm K}$, autoionizing states near the ionization threshold contribute to DR, making the rate sensitive to fine structure. Detailed DR calculations for astrophysical applications have been published for an increasing, but still incomplete, sample of $r$-process elements including Selenium ($Z=34$), Krypton ($Z=36$) \citep{Sterling&Witthoeft2011, Sterling2011}, Neodymium (Nd, $Z=60$) \citep{Hotokezaka+2021}, Rubidium ($Z=37$), Strontium ($Z=38$), Yttrium ($Z=39$), Zirconium ($Z=40$) \citep{Banerjee+2025}, Uranium ($Z=92$) \citep{Ferguson+2026}, and individual ion stages of Tellurium ($Z=52$) and Cerium ($Z=58$) \citep{Singh+2025}. 

We do an illustrative calculation of the ejecta temperature and ionization state for Nd to demonstrate the effects of particle transport and the ionization cascade, leaving a more self-consistent calculation including more species to future work. For simplicity, we assume that all ions are Nd, neglecting the coupling between ionization states of different elements via the shared pool of free electrons. We only include contributions to ionization and heating from $\beta$-particles, neglecting contributions from other decay products like $\alpha$-particles and fission fragments. We neglect other ionization sources: namely, thermal collisional ionization and photoionization. Thermal collisional ionization is found to always be subdominant in KN conditions. However, photoionization may be important at early times and for species with low ionization potentials \citep{Pognan+2023, Brethauer+2026}, so it should be included in future, more self-consistent calculations.

We use the temperature-dependent DR rate coefficients and cooling functions for Nd II, Nd III, and Nd IV from Figures~3 and 6 of \citet{Hotokezaka+2021} respectively. For Nd I and Nd V, we set the cooling function to that for Nd II and Nd IV respectively. For Nd I, we set the DR rate coefficient equal to $1/4$ that for Nd II. We use an analytic formula for the Case B RR rate coefficients from \citet{Axelrod1980}. Higher ion stages have higher DR rates and cool less efficiently. For all ion stages, the cooling function increases monotonically with temperature.

At each velocity coordinate in our fiducial ejecta model, we integrate Equations~\ref{eq:tempdot} and \ref{eq:xijdot}, taking the heating and per-species ionization rates from our Monte Carlo simulation. The equations are stiff and non-linear, so at each integration step, we iteratively compute the temperature and free electron density and take backward-Euler updates until convergence. The results are not sensitive to the initial conditions because, at early times, the thermal and ionization equilibrium timescales are short, so the ejecta is well described by a steady-state solution. Time-dependent effects only become important at late times as the equilibrium timescales grow \citep{Pognan+2022}.

\begin{figure*}
    \centering
    \includegraphics[width=0.8\linewidth]{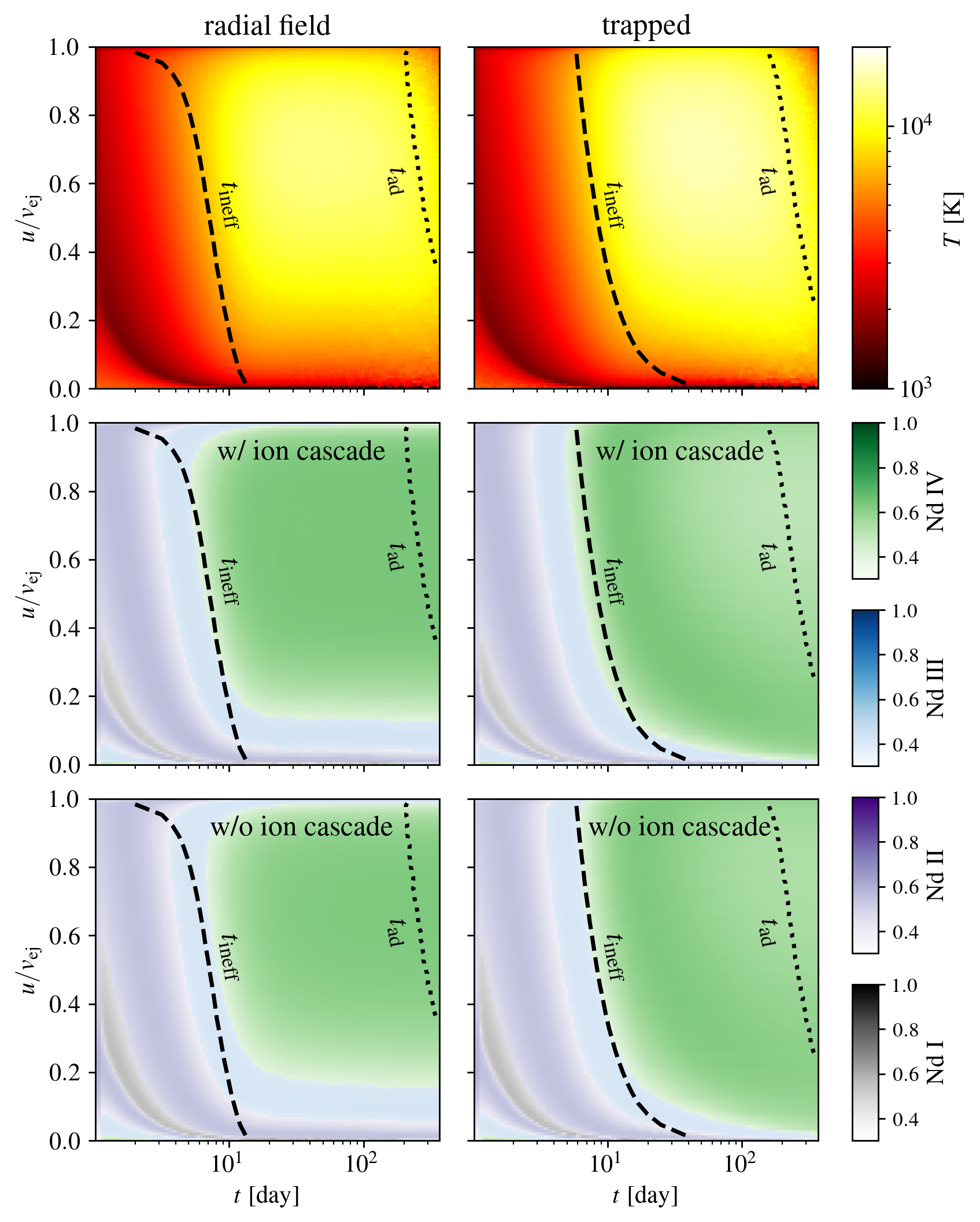}
    \caption{The temperature (first row) and ionization state (second row) of Neodymium $(Z=60)$ at each bin in velocity coordinate and time in our fiducial ejecta model in the radial magnetic field (left column) and trapped (right column) limits. Color represents the dominant ionization species, with intensity representing the fraction of that element that is in the dominant ionization state. Nd I, Nd II, Nd III, and Nd IV are represented by black, purple, blue, and green respectively. In each panel, we plot the time at which thermalization starts to become inefficient $t_{\rm ineff}$ (dashed black lines) and the time at which adiabatic cooling dominates $t_{\rm ad}$ (dotted black lines) at each velocity coordinate. In agreement with \citet{Brethauer+2026}, we find that the nebular ejecta has an inverted ionization structure due to less efficient recombination in the more diffuse outer ejecta and multiple ionization stages coexist at each location. Radial fields lead to a lower ionization state because particle escape lowers the ionization and heating rates. We also show the ionization state without the ionization cascade (third row; see Sec.~\ref{sec:cascade}), which lowers the ionization state throughout the ejecta. }
    \label{fig:ionstate}
\end{figure*}

In Figure~\ref{fig:ionstate}, we show the ejecta temperature and Nd ionization state for both transport scenarios, similar to Figure~3 of \citet{Brethauer+2026}. We also show the inefficiency time and the time when adiabatic cooling dominates at each velocity coordinate. In agreement with \citet{Brethauer+2026}, we find that the nebular ejecta has an inverted ionization structure due to less efficient recombination in the more diffuse outer ejecta and multiple ionization stages coexist at each location. Because our fiducial ejecta model ($M_{\rm ej} = 0.003~M_\odot$, $v_{\rm ej} = 0.2c$) produces lower densities than the ejecta model considered by \citet{Brethauer+2026} ($M_{\rm ej} = 0.03~M_\odot$, $v_{\rm ej} = 0.1 c$), our ejecta is more ionized than theirs, with Nd IV dominating after $\sim 10~{\rm days}$. Our results also deviate from theirs at early times when photoionization dominates over ionization from $\beta$-particles and photon absorption contributes significantly to heating.

The thermal evolution can be understood following the logic of \citet{Pognan+2022}. The line cooling rate scales as $\mathcal{C}_{\rm line} \propto \rho^2 \Lambda$. In the steady state and ignoring adiabatic cooling, we have $\mathcal{H} = \mathcal{C}_{\rm line}$, so the cooling function scales as $\Lambda \propto \rho^{-1} f_{\rm thm} Q_{\rm inj} \propto u^{\alpha} t^3 f_{\rm thm} Q_{\rm inj}$. Even though the energy injection rate and thermalization efficiency decrease over time, the density decreases faster due to homologous expansion, which requires the temperature to increase to support a higher equilibrium cooling function.

Initially, $f_{\rm thm}$ is approximately constant, so the cooling function increases rapidly $\Lambda \propto t^{1.7}$ and the ejecta heats up, reaching $\gtrsim 10000~{\rm K}$ after an inefficiency time. These temperatures are significantly hotter than models that neglect charged-particle ionization, for which the nebular-phase ejecta are mostly neutral and cool more efficiently \citep{Brethauer+2026, Fontes+2026}.

After a couple of inefficiency times, $f_{\rm thm}$ decreases rapidly, counterbalancing the falling density and causing the temperature to level off. At late times $\gtrsim 100~{\rm days}$, adiabatic cooling starts to contribute significantly and the outer ejecta cools. 

The same argument explains why the temperature peaks at intermediate velocity coordinates at a given time. In the inner ejecta, the equilibrium cooling function shifts to lower values due to the higher density, while in the outer ejecta, the function shifts to lower values due to less efficient thermalization.

As the temperature rises, the typical ionization stage increases, with the dominant species transitioning from Nd II/III to Nd IV. Radial fields promote particle escape, which lowers the ionization and heating rates, which also leads to lower ejecta temperatures and recombination rates. The net result is lower ionization stages throughout the ejecta. On average, the mean ion charge in corresponding bins is $0.24$ lower in the radial field limit than in the trapped limit. Transport physics affects the ionization state most strongly in the innermost and outermost ejecta, where it also has the strongest impact on thermalization efficiency (Fig.~\ref{fig:ethmu}). 

We also show the ionization state without the ionization cascade (see Sec.~\ref{sec:cascade}), which lowers the ionization state throughout the ejecta. On average, the mean ion charge in corresponding bins is $0.15$ lower when the ionization cascade is not included in the radial field limit. The effect is strongest in the inner ejecta, where more energy goes into the cascade. This calculation demonstrates that transport physics and the ionization cascade are necessary pieces for an accurate nebular-phase ionization structure.

\section{Conclusions}
\label{sec:conclusion}
We revisit the thermalization and transport of $\beta$-particles, which carry most of the decay energy that powers KNe light curves. For thermalization, we use evaluated atomic physics data to estimate per-species contributions to the stopping power, transport opacity, and ionization cross section, which we make available online (see the \hyperref[sec:data]{Data Availability statement}). For transport, we develop a fully relativistic framework for particle transport in a spherically symmetric, homologously expanding ejecta, considering two limiting magnetic-field geometries: particles moving through coherent radial magnetic fields and particles that are completely trapped. 

We study the impact of particle transport on thermalization across a range of ejecta models, varying the mass, velocity, ionization state, $r$-process abundance pattern, and density profile power-law. Overall, we find that charged-particle transport and realistic atomic physics significantly impact thermalization and ejecta ionization structure, and should be incorporated into the next generation of KN models. Our main findings are as follows:
\begin{itemize}

\item
\textit{$\beta$ decay rates and spectra}: Variations in the dominant radioactive nuclide produce shifts in the decay spectra and fluctuations in decay rate relative to the overall power-law trend, which can impact the thermalization efficiency.

\item
\textit{Stopping power and transport opacity}: At MeV energies, electron impact ionization (EII) dominates energy loss, while Coulomb interactions with ions dominate scattering. At lower energies reached during thermalization, energy loss and scattering are instead dominated by Coulomb interactions with free thermal electrons (M{\o}ller scattering). At higher ionization stages, EII contributions become weaker and M{\o}ller contributions become stronger.

\item
\textit{Secondary electrons}: A single $\beta$-particle can ionize thousands of atoms before thermalizing. These ionizations produce a large population of secondary electrons with typical energies $\sim 100~{\rm eV}$ and a high-energy tail. Ionizations by secondary electrons result in an ionization cascade that contributes significantly to the effective EII cross section, for which we provide an empirical fit (Eq.~\ref{eq:cascade_formula}) and include in our public data. The cascade contribution increases with incident electron energy, becoming similar to the primary contribution at 1 MeV.

\item
\textit{Charged particle transport}: As charged particles travel through the ejecta, they lose energy due to collisions and adiabatic expansion of the ejecta. The 1D transport equations for a charged particle in a coherent radial magnetic field are mathematically equivalent to those for an unmagnetized particle because the magnetic mirror force has the same effect as the rotation of the radial unit vector along a straight-line trajectory. At early times, particles thermalize locally in space and time. At intermediate times, particles persist in the ejecta and thermalize non-locally, increasingly escaping as scattering becomes less efficient. At late times, particles escape before most of their energy thermalizes. Particle escape becomes increasingly important relative to adiabatic losses during the diffusive transport phase, but stalls relative to adiabatic losses during the free-streaming transport phase.

\item
\textit{Thermalization efficiency}: Particle transport and escape reduce thermalization efficiency, especially in the innermost and outermost ejecta, making the efficiency more uniform across the ejecta compared to local-deposition models. In the trapped limit, particles persist in the ejecta for longer, so energy loss is driven by adiabatic expansion rather than particle escape. Thermalization efficiencies are sensitive to ejecta mass and velocity, as well as $r$-process abundance pattern and density power-law slope. However, they are less sensitive to ejecta ionization state, so energy deposition can be reasonably modeled without computing the ionization state self-consistently. At late times, results are broadly consistent with the \citet{Waxman+2019} model.

\item
\textit{Analytic prescription}: At each Lagrangian position in the ejecta, we describe the thermalization efficiency $f_{\rm thm}(t)$ with an analytic formula (Eq.~\ref{eq:anal_fit}) parameterized by $t_{\rm ineff}$, the time at which thermalization starts to become inefficient, and $n$, the asymptotic slope. Typical values are $t_{\rm ineff} \sim 10~{\rm days}$ and $n \simeq 1.6$, in broad agreement with previous work. We provide formulae for $t_{\rm ineff}$ and $n$ as a function of Lagrangian position and ejecta model parameters for use in future light curve calculations (Eqs.~\ref{eq:presfirst}--\ref{eq:preslast}, Table~\ref{tab:anal_fit}). This prescription is only approximate, so we recommend explicit charged-particle transport whenever feasible.

\item
\textit{Ejecta ionization structure}: As an illustrative example, we use the rate equations to compute the ionization structure of Neodymium for our fiducial ejecta model. In agreement with \citet{Brethauer+2026}, we find that the nebular ejecta has an inverted ionization structure due to less efficient recombination in the more diffuse outer ejecta and multiple ionization stages coexist at each location. Radial magnetic fields lead to a lower ionization state because particle escape lowers the ionization and heating rates. On the other hand, including ionization contributions from the ionization cascade raises the ionization state throughout the ejecta.

\end{itemize}

We focus on $\beta$-particles, but $\alpha$-particles and fission fragments also contribute significantly to heating and ionization, especially at late times, despite carrying a smaller fraction of the nuclear energy \citep{Barnes+2016}. The relevant microphysical processes have been analyzed by other works \citep{VandenBerg&Hotokezaka2026} and should be integrated into our analysis for a more complete description of thermalization and ionization. Our transport model could be applied to $\alpha$-particles and fission fragments in the radial field limit. However, at late times, these heavier particles cease to be well-magnetized and a more sophisticated model may be required. The peak-to-tail luminosity in the KN light curve may be sensitive to the relative contributions of different decay products, which can help constrain ejecta composition \citep{Barnes+2016}. 

Our illustrative calculation of the ejecta ionization structure should be coupled to a radiative transfer model (as done in \citealt{Brethauer+2026}) to include photoionization and ionization by recombination photons \citep{VandenBerg&Hotokezaka2026}. In addition, our calculation should be updated to include the ionization and excitation balance for all species, as done in \citet{Ricigliano+2025} for the study of $r$-process elements in supernova nebulae. The resulting radiation field and ejecta ionization structure could then be used to forward model observables.

We only consider particle transport for highly idealized magnetic-field geometries. Although we expect other magnetic field geometries to produce thermalization efficiencies in between our two limiting cases, there may be new behaviors that we do not capture due to spatially and temporally dependent magnetization, particle energization by diffusive shock acceleration, or collective plasma effects like the streaming instability. Future work should explore charged particle transport in realistic magnetic-field geometries and assess these possibilities using kinetic or hybrid approaches. 

We also expect thermalization to be affected by deviations from spherical symmetry, since realistic merger ejecta exhibit angular variations in density, composition, and velocity structure \citep{Kasen+2015, Wollaeger+2018, Bulla2019, Korobkin+2021}. Thermalization may be less efficient if transport can move decay products to more diffuse polar regions. These effects should also be explored in future work.

\section*{Acknowledgements}
ZA would like to thank Gian Luca Delzanno, Robert Ewart, Philipp Kempski, and Eliot Quataert for helpful conversations. The authors would like to thank Oleg Korobkin for data on the decay spectrum contributions from internal conversion electrons. The authors would like to thank Or Guttman, Ben Shenhar, and Eli Waxman for helpful conversations and for pointing out an error in the original version of this manuscript. We are grateful to our referee, whose time and attention improved the quality of this manuscript.

This material is based upon work supported by the U.S. Department of Energy, Office of Science, Office of Advanced Scientific Computing Research, Department of Energy Computational Science Graduate Fellowship under Award Number DE-SC0024386. 

This report was prepared as an account of work sponsored by an agency of the United States Government. Neither the United States Government nor any agency thereof, nor any of their employees, makes any warranty, express or implied, or assumes any legal liability or responsibility for the accuracy, completeness, or usefulness of any information, apparatus, product, or process disclosed, or represents that its use would not infringe privately owned rights. Reference herein to any specific commercial product, process, or service by trade name, trademark, manufacturer, or otherwise does not necessarily constitute or imply its endorsement, recommendation, or favoring by the United States Government or any agency thereof. The views and opinions of authors expressed herein do not necessarily state or reflect those of the United States Government or any agency thereof.

This work was supported by the U.S. Department of Energy through the Los Alamos National Laboratory and approved for public release under unlimited release tracking number LA-UR-26-25420. Los Alamos National Laboratory is operated by Triad National Security, LLC, for the National Nuclear Security Administration of U.S. Department of Energy (contract no. 89233218CNA000001).

The simulations presented in this article were performed on computational resources managed and supported by Princeton Research Computing, a consortium of groups including the Princeton Institute for Computational Science and Engineering (PICSciE) and the Office of Information Technology’s High Performance Computing Center and Visualization Laboratory at Princeton University.

\section*{Data Availability}
\label{sec:data}
Our atomic physics data, adapted from the EEDL database, are available online\footnote{\label{fn:data}\href{https://zandalman.com/publications/Andalman+2026b/}{zandalman.com/publications/Andalman+2026b/}}. These data include per-species contributions to the stopping, transport, and electron impact ionization cross sections, as well as combined cross sections using our strong $r$-process abundance pattern at $30~{\rm days}$ post-merger. For the stopping and transport cross sections, we separately tabulate the Mott and M{\o}ller contributions proportional to $\ln \Lambda$ and $\overline{q^2}$, allowing the data to be adapted to arbitrary values of these quantities. For the EII cross sections, we provide values including and excluding the contribution from the ionization cascade. The Monte Carlo data can be provided upon reasonable request to the corresponding author.

\bibliographystyle{mnras}
\bibliography{main}

\appendix
\section{PDF interpolation method}
\label{sec:algo}

The EEDL database provides differential EII cross sections in the secondary electron energy and differential Mott scattering cross sections in scattering angle at a set of discrete energies. To determine these differential cross sections at an arbitrary electron energy, we must interpolate a probability distribution function (PDF) from the given PDFs. Consider a tabulated dataset $\{(t_0, p_0(x)), (t_1, p_1(x)), \ldots, (t_N, p_N(x))\}$ where $p_i(x)$ is the PDF of $x$ for $t = t_i$. It would be convenient to interpolate between the tabulated PDFs using the standard interpolation formula
\begin{equation}
    p_t(x) = (1-\alpha) p_i(x) + \alpha p_{i+1}(x)\,,
\end{equation}
where $\alpha = (t - t_i) / (t_{i+1} - t_i)$ is the interpolation parameter. However, this produces undesirable results because $p_i(x)$ and $p_{i+1}(x)$ may be concentrated in different regions of the state space \citep{Bursal1996}. For example, one might incorrectly obtain a bimodal distribution by interpolating between two unimodal distributions. 

Instead, we interpolate the inverse cumulative distribution functions (CDF), or quantile functions. If $F_i(x)$ is the CDF of $x$ at $t = t_i$, then the interpolated inverse CDF is
\begin{equation}
    F_t^{-1}(q) = (1-\alpha) F_{i}^{-1}(q) + \alpha F_{i+1}^{-1} (q)\,.
    \label{eq:CDFinterp}
\end{equation}
The CDF is given by inverting $F_t^{-1}(q)$ and the PDF is given by differentiation $p_t(x) = \dd F_t(x) / \dd x$.

As a proof-of-concept, consider interpolating between two delta functions $\delta(x - x_0)$ and $\delta(x - x_1)$. The inverse CDFs are $F^{-1}_0(\xi) = x_0$ and $F^{-1}_1(\xi) = x_1$ respectively. Using Equation~\ref{eq:CDFinterp}, the interpolated inverse CDF is $x_i(\xi) = (1 - \alpha) x_0 + \alpha x_1$. This implies that the interpolated PDF is $\delta(x - x_i)$, which matches our intuitive expectation.

\section{Microphysical processes not modeled}
\label{sec:not_modeled}

\subsection{Sources of supra-thermal electrons}

When ionization leaves an atom with an inner-shell vacancy, the atom may undergo Auger-Meitner ionization, in which a higher-energy electron fills the vacancy and the liberated energy goes into unbinding a different electron in the same atom. However, these electrons are much less abundant than the secondary population discussed above because the inner shells only contribute a small fraction to the EII cross section and not every inner shell vacancy produces an Auger-Meitner electron, so we neglect their contribution to the ionization rate. This assumption is verified by \citet{VandenBerg&Hotokezaka2026}, who show that Auger-Meitner electrons contribute $\lesssim 10\%$ to the steady-state electron spectrum.

Internal conversion electrons can also be produced in some $\beta$ decays where the daughter nucleus ejects an electron during nuclear de-excitation. Using data from Oleg Korobkin (private communication), we compute the $\beta$ decay spectrum accounting for this population. However, we found that they had a negligible impact on the decay spectrum, so we do not include their contribution in our analysis.

Supra-thermal electrons can also be produced by $\gamma$-rays through Compton scattering of free thermal electrons and photoionization of bound electrons. However, the ejecta become transparent to $\gamma$-rays only a few days post-merger \citep{Barnes+2016}, so we neglect their contribution.

\subsection{Synchrotron losses}

The gyro-oscillations of an electron in a magnetic field with amplitude $B$ produce synchrotron radiation at a power given by
\begin{equation}
    P_{\rm sync} = \frac{2}{3} \frac{e^4 B^2}{m_{\rm e}^2 c^3} \gamma^2 \beta^2 (1 - \mu^2) \,,
    \label{eq:sync}
\end{equation}
where $B$ is the magnetic field amplitude and $\mu$ is the cosine of the pitch angle between the particle velocity and magnetic field direction. For a $1~{\rm MeV}$ electron in a $10^{-6}~{\rm G}$ field (see Sec.~\ref{sec:toy_model}), the synchrotron mass stopping power is $\sim 10^{-4}~{\rm eV\ cm^2/g}$, which is insignificant compared to other energy loss mechanisms, so we neglect synchrotron losses, similar to \citet{Barnes+2016}.

\subsection{Cherenkov radiation}
\label{sec:cher}

In a plasma, electrons with velocities faster than the phase velocity $v_{\varphi}$ excite electrostatic plasma oscillations along their trajectories, creating an electromagnetic wakefield analogous to the Cherenkov effect in optics \citep{Cohen1961}. An unmagnetized plasma can be modeled as a dispersive medium with relative dielectric constant $1 - (\omega_{\rm p}/\omega)^2$, where $\omega_{\rm p}$ is the plasma frequency, given by $\omega_{\rm p} = \sqrt{4\pi n_{\rm e} e^2 / m_{\rm e}}$.

By solving Maxwell's equations for an electron moving through the plasma, one can derive the power in Cherenkov radiation. In a na\"ive calculation, the Cherenkov power diverges. However, in practice, Landau damping inhibits the propagation of plasma waves when their phase velocity is close to the thermal velocity of free electrons, given by $v_{\rm thm} = \sqrt{3 k_{\rm B} T / (2 m_{\rm e})}$. 

Following \citet{Cohen1961}, we assume that the Cherenkov radiation spectrum is zero if $v_\varphi < \sqrt{2} v_{\rm thm}$. Under this assumption, the Cherenkov power is given by
\begin{equation}
    P_{\rm cher} = \frac{e^2 \omega_{\rm p}^2}{2 v} \ln\left[ \left( \frac{v}{v_{\rm thm}} \right)^2 - 1 \right] \,.
    \label{eq:cher}
\end{equation}
In the non-relativistic (NR) and ultra-relativistic (UR) limits, this gives a mass stopping power
\begin{equation}\begin{split}
	-\frac{1}{\rho} \eval{\dv{E}{x}}_{\rm cher} =\ & \frac{P_{\rm cher}}{\rho v}\\
	\simeq\ & \overline{q} \mu_{\rm mmw, 140}^{-1}
	\begin{cases}
		 280~E_{\rm MeV}^{-1} \Lambda_{\rm cher}~{\rm eV\ cm^2/g} & {\rm NR}\\
		 1100~\Lambda_{\rm cher, 0}~{\rm eV\ cm^2/g} & {\rm UR} \,,
	\end{cases}
\end{split}\end{equation}
where $\mu_{\rm mmw,140} = \mu_{\rm mmw} / 140$, $\Lambda_{\rm cher} = \ln(4/3) + \ln(E / k_{\rm B} T) \simeq 16.5 + \ln(E_{\rm MeV} / T_3)$, and $\Lambda_{\rm cher, 0} = \ln(2/3) + \ln(E_{\rm r} / k_{\rm B} T) \simeq 15.2 - \ln(T_3)$. At energies $\lesssim 10~{\rm keV}$, the Cherenkov mass stopping power is $\sim 10\%$ of the total mass stopping power (Fig.~\ref{fig:stop_kap}). We do not include Cherenkov losses in our model, because a consistent treatment requires matching energy loss contributions from collective effects and binary Coulomb collisions to avoid potential double-counting \citep{Brown+2005}.

This loss channel should be investigated in future work with a more sophisticated model, including the effect of additional modes that can be excited in a magnetized plasma, such as Whistler and ion cyclotron waves \citep{McKenzie1967}. These effects have been explored by previous work in the context of pulsed electron beams in the Earth's magnetosphere \citep{Harker&Banks1984, Farrell&Goertz1990, Delzanno&Roytershteyn2019}, but not for individual supra-thermal electrons.

\section{Thermalization efficiency fit parameters and residuals}
\label{app:fit}

In this section, we provide tables for the inefficiency time and asymptotic slope for all ejecta models, determined by fitting Equation~\ref{eq:anal_fit} to the thermalization efficiency integrated over velocity coordinate. In Tables~\ref{tab:mejvej} and \ref{tab:mejvej2}, we show results for our series in ejecta mass and velocity with density power-law exponents $\alpha=1$ and $\alpha=2$ respectively. In Table~\ref{tab:ir}, we show results for our series in ejecta ionization stage and $r$-process abundance pattern. 

\begin{figure}
    \centering
    \includegraphics[width=1.0\linewidth]{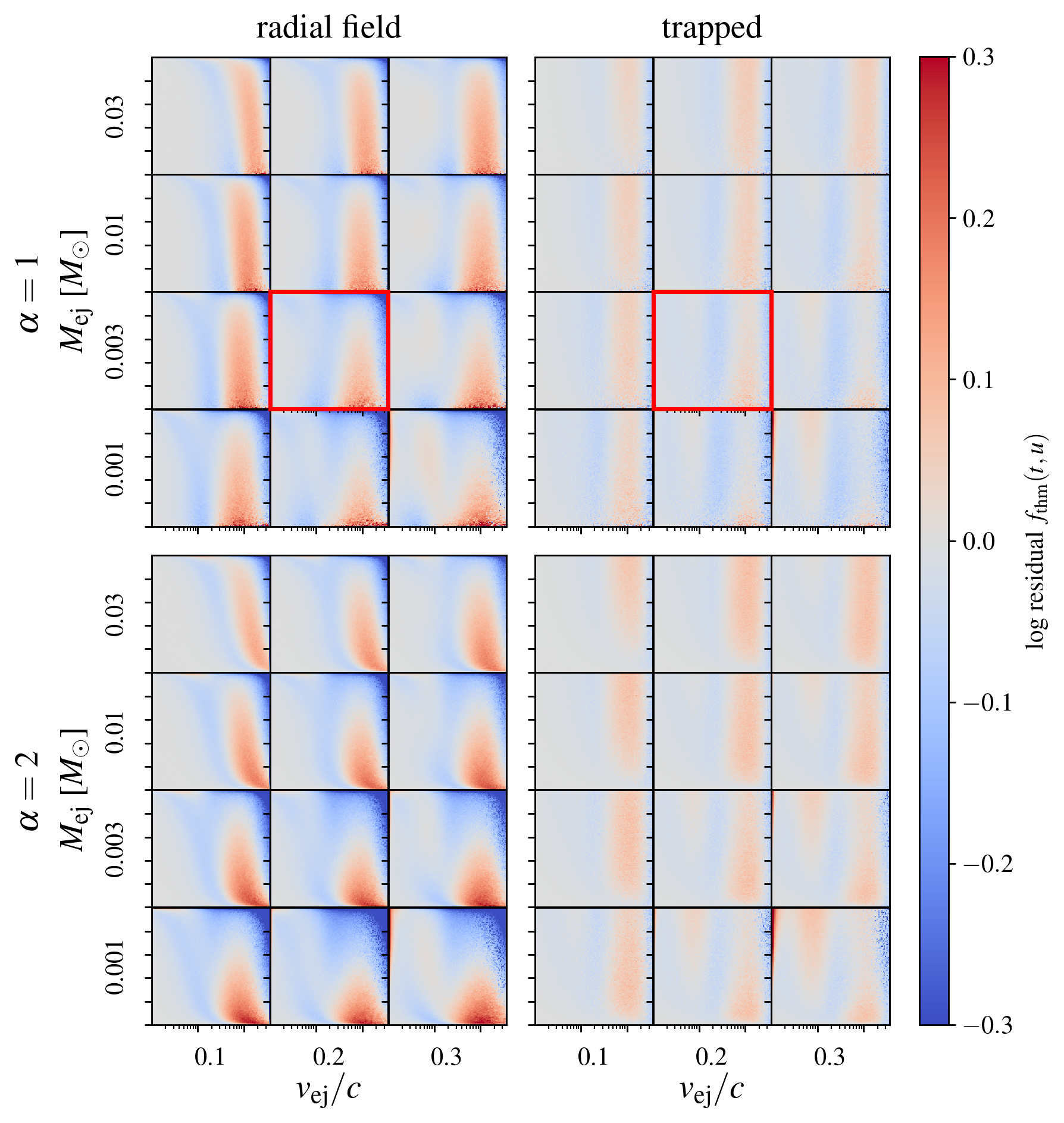}
    \caption{Same as Figure~\ref{fig:res}, but with each panel representing a different ejecta model with $x$- and $y$-axes identical to Figure~\ref{fig:res}. The top and bottom panel groups show models with density power-law slopes $\alpha=1$ and $\alpha=2$ respectively. The left and right panel groups show models in the radial field and trapped limits respectively. Within each panel group, rows correspond to different ejecta masses and columns correspond to different ejecta velocities. The panels corresponding to the fiducial model are outlined in red. In the trapped limit, all models have logarithmic residuals $\lesssim 0.1~{\rm dex}$ at all locations and times. In the radial field limit, the largest residuals $\sim 0.3~{\rm dex}$ occur in the innermost and outermost ejecta at late times, mostly because we approximate the asymptotic slope as independent of velocity coordinate. These results indicate that our prescription is a reasonable approximation across a range of ejecta conditions.}
    \label{fig:res_all}
\end{figure}

We also provide logarithmic residuals for our analytic prescription for the spatially-dependent thermalization efficiency described in Section~\ref{sec:analytic_prescription}. These are shown in Figure~\ref{fig:res_all}, similar to the residuals shown in Figure~\ref{fig:res} for the fiducial ejecta model.

\begin{table}
    \centering
    \caption{The inefficiency time $t_{\rm ineff}$ and asymptotic slope $n$ for ejecta models in our series in ejecta mass and velocity with density power-law slope $\alpha=1$, determined by fitting Equation~\ref{eq:anal_fit} to the thermalization efficiency integrated over velocity coordinate. We show values in the radial magnetic field (superscript rad) and trapped (superscript trp) limits. The fiducial model is bolded. }
    \label{tab:mejvej}
    \begin{tabular}{cccccc}
        \hline
        $M_{\rm ej}$ [$M_\odot$] & $v_{\rm ej}/c$ & $t_{\rm ineff}^{({\rm rad})}$ [day] & $n^{({\rm rad})}$ & $t_{\rm ineff}^{({\rm trp})}$ [day] & $n^{({\rm trp})}$ \\
        \hline
        0.001 & 0.1 & 7.31 & 1.96 & 12.2 & 1.56 \\
        0.003 & 0.1 & 12.2 & 1.95 & 20.0 & 1.54 \\
        0.01 & 0.1 & 21.5 & 1.94 & 34.8 & 1.49 \\
        0.03 & 0.1 & 36.9 & 1.93 & 60.1 & 1.45 \\
        0.001 & 0.2 & 3.31 & 1.87 & 4.28 & 1.52 \\
        \textbf{0.003} & \textbf{0.2} & \textbf{5.78} & \textbf{1.89} & \textbf{7.57} & \textbf{1.55} \\
        0.01 & 0.2 & 10.3 & 1.90 & 13.4 & 1.56 \\
        0.03 & 0.2 & 17.1 & 1.88 & 22.0 & 1.53 \\
        0.001 & 0.3 & 1.89 & 1.77 & 2.13 & 1.47 \\
        0.003 & 0.3 & 3.44 & 1.81 & 3.97 & 1.51 \\
        0.01 & 0.3 & 6.37 & 1.83 & 7.44 & 1.55 \\
        0.03 & 0.3 & 10.8 & 1.84 & 12.6 & 1.56 \\
        \hline
    \end{tabular}
\end{table}

\begin{table}
    \centering
    \caption{Same as Table~\ref{tab:mejvej}, but with density power-law slope $\alpha=2$.}
    \label{tab:mejvej2}
    \begin{tabular}{cccccc}
        \hline
        $M_{\rm ej}$ [$M_\odot$] & $v_{\rm ej}/c$ & $t_{\rm ineff}^{({\rm rad})}$ [day] & $n^{({\rm rad})}$ & $t_{\rm ineff}^{({\rm trp})}$ [day] & $n^{({\rm trp})}$ \\
        \hline
        0.001 & 0.1 & 9.61 & 1.73 & 16.1 & 1.20 \\
        0.003 & 0.1 & 16.1 & 1.71 & 27.3 & 1.21 \\
        0.01 & 0.1 & 28.5 & 1.68 & 48.8 & 1.20 \\
        0.03 & 0.1 & 48.7 & 1.65 & 85.3 & 1.20 \\
        0.001 & 0.2 & 4.63 & 1.70 & 5.62 & 1.17 \\
        0.003 & 0.2 & 7.80 & 1.68 & 9.87 & 1.19 \\
        0.01 & 0.2 & 13.7 & 1.67 & 17.8 & 1.21 \\
        0.03 & 0.2 & 22.9 & 1.63 & 30.2 & 1.21 \\
        0.001 & 0.3 & 2.88 & 1.66 & 2.87 & 1.14 \\
        0.003 & 0.3 & 4.93 & 1.65 & 5.23 & 1.17 \\
        0.01 & 0.3 & 8.73 & 1.64 & 9.73 & 1.19 \\
        0.03 & 0.3 & 14.5 & 1.61 & 16.7 & 1.20 \\
        \hline
    \end{tabular}
\end{table}

\begin{table}
    \centering
    \caption{Same as Table~\ref{tab:mejvej}, but for our series in ejecta ionization state and $r$-process abundance pattern. The fiducial model is bolded.}
    \label{tab:ir}
    \begin{tabular}{cccccc}
        \hline
        ion stage & $r$-process & $t_{\rm ineff}^{({\rm rad})}$ [day] & $n^{({\rm rad})}$ & $t_{\rm ineff}^{({\rm trp})}$ [day] & $n^{({\rm trp})}$ \\
        \hline
        II & weak & 6.98 & 1.82 & 9.38 & 1.47 \\
        III & weak & 7.3 & 1.78 & 9.98 & 1.42 \\
        IV & weak & 7.78 & 1.75 & 10.7 & 1.39 \\
        II & medium & 6.36 & 1.90 & 13.3 & 1.61 \\
        III & medium & 6.12 & 1.84 & 13.1 & 1.56 \\
        IV & medium & 6.14 & 1.82 & 13.5 & 1.54 \\
        \textbf{II} & \textbf{strong} & \textbf{5.78} & \textbf{1.89} & \textbf{7.57} & \textbf{1.55} \\
        III & strong & 5.63 & 1.84 & 7.40 & 1.51 \\
        IV & strong & 5.6 & 1.82 & 7.39 & 1.48 \\
        \hline
    \end{tabular}
\end{table}

\bsp	
\label{lastpage}
\end{document}